%% file: 21fupa.tex
\documentclass[sigconf,nonacm]{acmart}

\input{./21fupa_main_wsdm/21fupa_pre}

\AtBeginDocument{%
  \providecommand\BibTeX{{%
    \normalfont B\kern-0.5em{\scshape i\kern-0.25em b}\kern-0.8em\TeX}}}

\setcopyright{acmcopyright}
\copyrightyear{2018}
\acmYear{2018}
\acmDOI{10.1145/1122445.1122456}

\acmConference[Woodstock '18]{Woodstock '18: ACM Symposium on Neural
  Gaze Detection}{June 03--05, 2018}{Woodstock, NY}
\acmBooktitle{Woodstock '18: ACM Symposium on Neural Gaze Detection,
  June 03--05, 2018, Woodstock, NY}
\acmPrice{15.00}
\acmISBN{978-1-4503-XXXX-X/18/06}

\begin{document}

\title[]{On Designing a Two-stage Auction for Online Advertising}
\input{./21fupa_main_wsdm/21fupa_author}
\begin{abstract}
  \input{./21fupa_main_wsdm/21fupa_abs.tex}

\end{abstract}




\maketitle

\input{./21fupa_main_wsdm/21fupa_doc}
\bibliographystyle{ACM-Reference-Format}
\bibliography{sample-base.bib, 21fupa.bib}

\appendix
\input{./21fupa_main_wsdm/21fupa_appendix}









\end{document}

%% file: 21fupa_main_wsdm/21fupa_pre.tex
\usepackage{subfig}
\usepackage{multirow}
\usepackage{verbatim}
\usepackage{bbm}

%% file: 21fupa_main_wsdm/21fupa_author.tex
\author{Yiqing Wang$^1$, Xiangyu Liu$^2$, Zhenzhe Zheng$^1$, Zhilin Zhang$^2$, Miao Xu$^2$, Chuan Yu$^2$ and Fan Wu$^1$}
\renewcommand{\shortauthors}{
}
\affiliation{
\institution{$^1$Shanghai Jiao Tong University, $^2$Alibaba Group}
\and
\{wangyiqing\_2015, zhengzhenzhe\}@sjtu.edu.cn, fwu@cs.sjtu.edu.cn
\and
\{qilin.lxy, zhangzhilin.pt, xumiao.xm, yuchuan.yc\}@alibaba-inc.com
\country{}
}

%% file: 21fupa_main_wsdm/21fupa_abs.tex

For the scalability of industrial online advertising systems, a two-stage auction architecture is widely used to enable efficient ad allocation on a large set of corpus within a limited response time. The current deployed two-stage ad auction usually retrieves an ad subset by a coarse ad quality metric in a pre-auction stage, and then determines the auction outcome by a refined metric in the subsequent stage. However, this simple and greedy solution suffers from serious performance degradation, as it regards the decision in each stage separately, leading to an improper ad selection metric for the pre-auction stage. 
In this work, we explicitly investigate the relation between the coarse and refined ad quality metrics, and design a two-stage ad auction by taking the decision interaction between the two stages into account. We decouple the design of the two-stage auction by solving a stochastic subset selection problem in the pre-auction stage and conducting a general second price (GSP) auction in the second stage. We demonstrate that this decouple still preserves the \emph{incentive compatibility} of the auction mechanism. As the proposed formulation of the pre-auction stage is an NP-hard problem, we propose a scalable  approximation solution by defining a new subset selection metric, namely \emph{Pre-Auction Score (PAS)}. Experiment results on both public and industrial dataset demonstrate the significant improvement on social welfare and revenue of the proposed two-stage ad auction, than the intuitive greedy two-stage auction and other baselines.

%% file: 21fupa_main_wsdm/21fupa_doc.tex
\newcommand*{\dif}{\mathop{}\!\mathrm{d}} 
\newcommand*{\E}{\mathbb{E}}
\def\bs#1{\boldsymbol{#1}} 

\newcommand*{\setA}{\mathcal{A}}
\newcommand*{\stk}{\mathtt{SumTopK}}
\newcommand*{\setD}{\mathcal{D}}
\newcommand*{\sx}{\mathcal{S}}
\newcommand*{\clk}{\mathrm{CLK}}
\newcommand*{\M}{\mathtt{M}}
\def\ie{\emph{i.e.}\,}
\def\eg{\emph{e.g.}\,}
\def\etal{\emph{et al.}\,}


\section{Introduction}

Online advertising is the major sources of revenue for Internet industry~\cite{edelman2007internet}. 
Modern online advertising platforms conduct ad allocation by running ad auction mechanisms in real-time. 
To achieve effectiveness and efficiency in ad allocation, the ad auction mechanisms jointly consider the bids from advertisers as well as the quality of displaying ads to users. 
For example, in the celebrated GSP auction~\cite{varian2007position,edelman2007internet}, the ad allocation decisions are made based on the metric of bid multiplying the ad quality. 
The ad quality is measured by the potential actions (\eg, click and purchase) of users on the displayed ads, such as click through rate (CTR)~\cite{he2014practical,cheng2016wide,zhou2018deep} and 
conversion rate (CVR)~\cite{lee2012estimating}, which can be estimated by learning models over the rich features of users and ads and the abundant data of user-ad interaction\footnote{Without loss of generality, we regard CTR as the ad quality in this work.}. 
Thus, the performance of ad allocation depends not only on the rule of auction mechanism, but also on the accuracy of learning models. 


In practice, direct implementation of such a one-stage ad auction mechanism faces scalability issues, and a two-stage auction architecture is used to make a trade-off between the system scalability and ad allocation performance. 
The one-stage ad auction mechanism must decide the auction outcome on a set of thousands of ads within tens of milliseconds~\cite{he2014practical, zhu2017optimized}. 
However, the sophisticated learning models, such as Wide\&Deep \cite{cheng2016wide} and DIN \cite{zhou2018deep}, can not complete the CTR estimations for all the candidate ads within a limited response time. 
To relieve this dilemma between performance and scalability, we turn to the two-stage architecture for online ad auction, 
which is also adopted for large-scale online recommenders~\cite{covington2016deep, eksombatchai2018pixie}.
An intuitive and greedy design of such a two-stage ad auction is illustrated in Figure \ref{fig:problem-gdy}: in the first stage, called pre-auction stage, we rank the full set of $N$ ads $\setA_N$, and select a subset of $M$ ads $\setA_M$ by $bid$ multiplying $\widetilde{ctr}$ from a fast and coarse ctr estimator $\M^c$; then in the second stage, called auction stage, we determine the final ad allocation results
by evaluating only on the selected ad subset 
via $bid$ multiplying $ctr$ from a sophisticated and refined ctr estimator $\M^r$.
Although this greedy two-stage auction mechanism has been widely deployed in industry~\cite{he2014practical,wang2020cold}, it suffers serious loss on ad allocation performance. 
Due to the ability gap between the two $ctr$ estimators, the coarse and the refined $ctr$s sometimes differ greatly for the same $\langle\text{ad,user}\rangle$ pair. 
Some ads with high refined $ctr$ but low coarse $ctr$ could be filtered out by the first stage, losing the chance of entering the second stage and winning an ad slot. 

\input{./21fupa_main_wsdm/21fupa_fig_problem.tex}


This greedy design reflects a common misunderstanding in applying the two-stage architecture for online advertising.  
While optimizing ad allocation performance for the two-stage auction, just regarding each stage as a separate optimization problem, \eg, conducting the individual GSP auction with the same metric of $bid\times ctr$ for ad selection in each stage, would suffer performance degradation. 
The auction designer should consider the interaction between the two stages (\ie, the second stage will refine the estimation and selection over the subset delivered by the first stage), and figure out proper selection metrics for each stage. 


In this work, we focus on the problem of designing a two-stage auction for online advertising, jointly considering the scalability of large-scale online systems and the ad allocation performance guarantee. 
There are two major challenges of this work. 
The first challenge is to characterize the conditions of the two-stage ad auction to satisfy the economic properties of \emph{incentive compatible (IC)}, \ie, the advertisers are encouraged to report their truthful value to a user click as bid, and \emph{individual rational (IR)}, \ie, the utility for an advertiser is always non-negative. 
The second challenge is the large decision space and then the high computational complexity of designing the two-stage auctions, \ie, how can we decouple the auction design of the two stages and propose specific auctions for each stage, so that the whole two-stage auction has performance guarantee within limited response time. 
Although there exist many works for designing ad auction mechanisms to optimize variance of performance metrics such as social welfare and revenue~\cite{lahaie2007revenue,varian2007position}, none of them considered the two-stage auction architecture. 
At the same time, although the two-stage recommender system gradually attracts attention recently~\cite{covington2016deep, eksombatchai2018pixie,hron2021component}, the two-stage problem of online advertising differs from that of recommender systems because there are payment transfer as well as the requirement of economic properties of IC and IR in ad auctions, while neither of these two issues appears in recommender systems.


The intuition behind our proposed solution for the two-stage ad auction is as follows. 
First, we specify the utility model for advertisers as \emph{value maximizers}, a proper model to capture the objectives of advertisers in online advertising~\cite{wilkens2017gsp, liu2021neural}, and then obtain the characteristics of the two-stage auction to be IC and IR under this model. 
Then, to reduce the design space, we fix the second stage auction as the celebrated general second price (GSP) auction, and demonstrate that this decoupling preserves the  optimality of performance and the properties of IC and IR for the two-stage auction. 
Next, we focus on the design in the first stage, \ie, the pre-auction stage, and formulate it as a stochastic subset selection problem, which is to select $M$ stochastic elements to maximize the expected sum of the realized top $K$ elements when we regard the unrealized $ctr$s of ads in the pre-auction stage as random variables. 
However, this subset selection problem is a submodular maximization with cardinality constraints, which can be proved to be NP-hard. 
To derive a scalable pre-auction, we turn to a closely related problem of selecting $M$ stochastic elements which maximizes the expected recall on the realized top $K$ elements, based on which we can define a new ad-wise selection metric, namely \emph{Pre-Auction Score (PAS)}. 
We further design a learning-based implementation of this PAS metric, which can be trained with the supervision by the refined ctr estimator $\M^r$.  
As PAS is a more proper metric for the pre-auction stage, our solution outperforms the widely adopted greedy design. 
We illustrate the detailed procedure of our two-stage auction in Figure \ref{fig:problem-pas}.

We summarize the main contributions of this work as follows. 
\begin{itemize}
\item We define a new mechanism design problem of two-stage ad auction design for online advertising, which jointly considers system scalability and performance guarantee. 
We also demonstrate the  performance degradation
of a widely adopted two-stage ad auction, which ignores the interaction between the optimization problems in the two stages. 
\item We propose a solution for designing a two-stage ad auction for social welfare maximization. We decouple the two-stage auction as a stochastic subset selection problem in the pre-aucton stage and the GSP auction in the second stage. 
We propose a new ad-wise selection metric Pre-Auction Score (PAS) to solve the subset selection. 
The proposed two-stage ad auction still satisfies the properties of IC and IR.
\item Extensive experiments on both public and industrial data demonstrate the effectiveness of our solution.  
On the industrial data, our solution outperforms the greedy two-stage auction by +4.35\% on social welfare and +4.59\% on revenue. 
\end{itemize}


\section{Preliminaries}

In this section, we describe the ad auction model, the utility model of advertisers, and the desired economic properties of ad auctions. 


For an online ad platform, a page view request from a user triggers an ad auction, where a set of $N$ advertisers $\setA_N=\{1,\ldots,N\}$ compete for $K$ ad slots on the page. 
In an ad auction, each advertiser $i\in\setA_N$ submits a bid $b_i$ based on her \emph{private} value $v_i$, which measures the potential benefits extracted from the user click on the ad. 
Besides the bids of advertisers $\bs{b}=(b_1,\ldots,b_N)$, the auction mechanism also depends on the advertiser/ad features $\bs{a}=(a_1,\ldots,a_N)$ and the user features $u$.
Here, the the advertiser/ad features $a_i$ could be ad id, category id, 
and etc. 
The user features $u$ could be user id, click histories, and etc. 
The features of ads and users are applied to evaluate the quality of displaying a certain ad to a specific user. 
The auction mechanism $\langle \bs{x},\bs{p} \rangle$ determines the ad allocation outcomes by the allocation scheme $\bs{x}$ and the prices of the $K$ ad slots by the payment rule $\bs{p}$. 
The allocation scheme $x_i(\bs{b},\bs{a}, u)=k$ represents that the ad $i$ wins the $k$-th highest ad slot, or loses the auction for $k=0$; and the payment rule $p_i(\bs{b},\bs{a},u)$ is the price that the ad $i$ needs to pay if it is displayed and clicked. 

We next describe the utility model of advertisers. 
The advertisers would like to maximize the advertising performance of their products, only requiring the costs to satisfy certain constraints, such as budget constraint, cost-per-click constraint and return-of-investment constraint \cite{wu2018budget, yang2019bid}. 
Following the industrial observations from \cite{liu2021neural,balseiro2021landscape}, \emph{value maximizer} model \cite{wilkens2017gsp} captures such an optimization objective of the advertisers, while the traditional \emph{quasi-linear utility maximizer} model~\cite{myerson1981optimal} (\ie, each advertiser $i$ maximizes $v_i-p_i$) is no longer suitable in this scenario. 
\begin{definition}{(\emph{Value Maximizer}~\cite{wilkens2017gsp})}
\label{def:value-maximizer}
An advertiser $i$ is a value maximizer if she prefers the auction outcomes with a higher ad slot while keeping the payment satisfy the constraint $p_i\leq v_i$; for the same ad slot, she prefers a lower payment $p_i$. 
\end{definition}


In an ad auction, the advertisers might strategically misreport their values, \ie, bidding $b_i\neq v_i$, to manipulate the auction outcomes for their own interests. 
To avoid this kind of behavior, the ad auction mechanism needs to satisfy the following economic properties: 
\begin{itemize}
\item \emph{Incentive Compatibility (IC)}: truthfully reporting the private value, \ie, $b_i=v_i$, is the best strategy for each advertiser $i$. 
\item \emph{Individual Rationality (IR)}: the payment of advertiser $i$ would not exceed the reported value, \ie, advertiser $i$ pays $p_i\leq b_i$ if ad $i$ is displayed and clicked; or pays nothing, otherwise. 
\end{itemize}
With these two properties, advertisers do not need to spend efforts in computing bidding strategy, and are encouraged to participate in the auctions with no risk of deficit. 
The online platform also obtains the truthful and reliable advertisers' values. 


For the advertisers with the utility model as a value maximizer, it has been proven in \cite{wilkens2017gsp} that an auction mechanism satisfies IC and IR if the following two conditions are satisfied: 
\begin{itemize}
\item \textbf{Monotonicity}: An advertiser would win the same or a higher ad slot if she reports a higher bid; 
\item \textbf{Critical price}: The payment for a winning advertiser is the minimum bid that she needs to maintain the same ad slot. 
\end{itemize}


The goal of the ad auction mechanism considered in this work is to maximize the expected \emph{social welfare}, which is the sum of the expected click values of displayed ads with respective to user's stochastic click behaviors. 
Social welfare is a crucial metric for online advertising, as it measures the efficiency on matching advertisers and users, and is also the upper bound of the revenue which is the sum of the total payments of the ad platform. 

\section{Problem Formulation}

\subsection{One-stage Ad Auction}


One of the most widely used ad auction mechanisms is GSP auction~\cite{varian2007position,edelman2007internet}.
Given each advertiser $i$'s bid $b_i$ for a user click along with the user's $ctr_i$ to the ad $i$, the GSP auction ranks all the ads with their expected click values of display, \ie, $b_i\times ctr_i$, and allocates the $K$ ad slots from the highest to the lowest following this rank. 
The payment for the winning ad at the slot $k\leq K$ is $b_{(k+1)} \times ctr_{(k+1)} / b_{(k)}$, where the subscript $(k)$ denotes the ad with the $k$-th highest expected click value. 
GSP auction is IC and IR for value-maximizing advertisers as it satisfies the conditions of monotonicity and critical price mentioned above \cite{wilkens2017gsp}. 
Thus, in GSP auction we have $b_i=v_i$ for each ad $i$. 
When there is an unbiased $ctr$ estimator\footnote{Various calibration algorithms can be applied to augment the basic $ctr$ estimator to further reduce the bias~\cite{deng2020calibrating}.}, the GSP auction can maximize the expected social welfare~\cite{varian2007position}, \ie, the total expected click value of the $K$ winning ads:
$\stk( \{v_i\times ctr_i\}_{i\in\setA} ) = \stk( \{b_i\times ctr_i\}_{i\in\setA } )$.
Here, $\stk(\sx)$
is a set function which outputs the sum of the largest $K$ elements in the set $\sx$. 


\subsection{CTR Estimator in Ad Auction}


The performance of ad auctions largely depends on the accuracy of ctr estimators. 
There are various kinds of machine learning models developed in the literature to estimate the ctr in different scenarios. 
We classify these models into two categories: the coarse but fast $ctr$ estimator denoted by $\M^c$ and the refined but heavy $ctr$ estimator denoted by $\M^r$.
The coarse $ctr$ estimator $\M^c=\widetilde{ctr}(\tilde{a}_i,\tilde{u})$ uses light-weight learning models~\cite{he2014practical,wang2020cold}, and only leverages partial ad features $\tilde{a}_i$ and partial user features $\tilde{u}$. 
The refined $ctr$s estimator $\M^r=ctr(a_i,u)$ can be sophisticated learning models~\cite{zhou2018deep,zhou2019deep}, and effectively leverages full ad features $a_i$ and user features $u$. 
For example, the full user features $u$ can be the user profiles along with a long histories of user's behaviors~\cite{zhou2019deep}, while the partial user features $\tilde{u}$ are just some basic user profiles. 
The refined estimator $\M^r$ could have a complex neural network architecture, such as sequence modeling components, to produce rich user-ad cross features \cite{zhou2018deep,zhou2019deep}. 
In contrast, the coarse estimator $\M^c$ might simply follow a two-tower architecture \cite{he2014practical} or an embedding layer followed by fully connected layers \cite{wang2020cold}. 
With these differences, the $\M^r$ estimator consumes longer inference time but produces more accurate $ctr$ than the $\M^c$ estimator does. 


We next investigate the relation between the coarse estimator $\M^c$ and the refined estimator $\M^r$.
Suppose the models of both $\M^c$ and $\M^r$ are sufficiently trained with the same data set $\setD$, 
which is independently and identically sampled from true distribution of $\langle\text{user,ad}\rangle$ pairs. 
For the input of the full feature $(a_i,u)$ with the corresponding partial feature $(\tilde{a}_i,\tilde{u})$, the outputs of $\M^c$ and $\M^r$ converge as below:
\begin{equation}
\label{eq:ctr-converge}
\begin{aligned}
& ctr(a_i,u) = | \setD^+(a_i,u) | / | \setD(a_i,u) |, \\
& \widetilde{ctr}(\tilde{a}_i,\tilde{u}) 
= \frac{ | \setD^+(\tilde{a}_i,\tilde{u}) | }{ | \setD(\tilde{a}_i,\tilde{u}) | } 
= \sum_{a_i|\tilde{a}_i,u|\tilde{u}} ctr(a_i,u) \frac{ | \setD(a_i,u) | }{ | \setD(\tilde{a}_i,\tilde{u}) | },
\end{aligned}
\end{equation}
where $\setD(a,u)\subseteq\setD$ is the subset of samples whose partial or full features are restricted to $(a,u)$, and $\setD^+$ is the subset of positive samples, \ie, the clicked samples.
Since $\frac{ | \setD(a_i,u) | }{ | \setD(\tilde{a}_i,\tilde{u}) | } \approx \Pr[a_i,u|\tilde{a}_i,\tilde{u}]$, the relation between $\M^r$ and $\M^c$ on input $(a_i,u)$ can be approximately expressed as 
\begin{equation}
\label{eq:ctr-relation}
\begin{array}{rl}
\widetilde{ctr}(\tilde{a}_i,\tilde{u}) & = \sum_{a_i|\tilde{a}_i,u|\tilde{u}} \,\, ctr(a_i,u)\times \Pr[a_i,u|\tilde{a}_i,\tilde{u}] \\
& = \mathbb{E}[ ctr(a_i,u) | \tilde{a}_i,\tilde{u} ].
\end{array}
\end{equation} 
According to this relation, serving the one-stage GSP auction with $\M^c$ and $\M^r$ estimators 
results in expected social welfare in (\ref{eq:gsp-sw-c}) and (\ref{eq:gsp-sw-r}), respectively:
\begin{subequations}
\begin{align}
& \mathbb{E}[ \stk( \{ b_i \times \widetilde{ctr}(\tilde{a}_i,\tilde{u}) \}_{i\in\setA_N} ) ] 
\label{eq:gsp-sw-c} \\
= & \mathbb{E}[ \stk( \{ \mathbb{E}[b_i \times ctr(a_i,u)|\tilde{a}_i,\tilde{u}] \}_{i\in\setA_N} ) ] 
\notag \\
\leq & \mathbb{E}[ \stk( \{ b_i \times ctr(a_i,u) \}_{i\in\setA_N} ) ].
\label{eq:gsp-sw-r} 
\end{align}      
\end{subequations}
The inequality in (\ref{eq:gsp-sw-r}) is due to Jensen's inequality for convex function and the fact that $\stk$ here can be regarded as 
a convex function over $\mathbb{R}^N$. 
We see that the one-stage GSP auction with the refined $\M^r$ estimator achieves strictly higher expected social welfare than that with the coarse estimator $\M^c$. 

\subsection{Two-stage Ad Auction}


Due to the scalability requirement of determining auction allocation and payment over thousands of ads within tens of milliseconds, we are unable to implement the one-stage GSP auction mentioned above, 
because applying the refined $ctr$ estimator $\M^r$ to the total set of ads $\setA_N$ exceeds the decision latency with limited computational resources.
To make a trade-off between the optimality of social welfare and the timely response time, the architecture of two-stage ad auction is widely used in practice~\cite{he2014practical,wang2020cold}. 
Specifically, the first stage, called the pre-auction stage associated with an allocation scheme $\bs{x}^p$, selects a subset of $M<N$ ads, \ie, $\bs{x}^p(\setA_N) = \setA_M \subsetneq \setA_N$. 
With the input of the selected ads $\setA_M$ from the first stage, the second stage, called the auction stage, determines the final allocation $\bs{x}^a$ and payment $\bs{p}^a$. 
The pre-auction stage uses coarse but fast machine learning models with partial ad $\bs{\tilde{a}}$ and user features $\bs{\tilde{u}}$ on the large set $\setA_N$; while the second auction stage can apply more advanced and accurate $ctr$ estimators with full ad and user features on a relatively small set $\setA_M$. 
In the next section, we demonstrate that without a careful design of the two-stage auction, we may suffer from a performance degradation. 


Compared with the one-stage auction design, two challenges immediately emerge for the two-stage auction design. 
First, how to guarantee the economic properties of IC and IR in a two-stage  auction.  
Second, considering that the searching space for jointly designing auctions in the two stages is huge, how to decouple the design of the two-stage auction and still guarantee the ultimate auction performance. 
We recall that (i) GSP auction is IC and IR for value maximizer advertisers, 
and (ii) GSP auction can maximize the expected social welfare when there is a refined ctr estimator. 
Due to these two advantages of GSP auction, we can fix the second stage as GSP auction, which introduces neither the violence of IC and IR nor the loss of optimality for expected social welfare. 


We focus on the auction design in the first stage, \ie, the pre-auction stage, in this work. Firstly, we coordinate the pre-auction and the GSP auction to satisfy the two conditions (monotonicity and critical price) for IC and IR properties. 
Since there is no payment in the pre-auction stage, we only need to guarantee that the allocation scheme $\bs{x}^p$ satisfies monotonicity, \ie, $x^p_i(\bs{b},\tilde{\bs{a}}, \tilde{u})$ is monotone increasing with respective to $b_i$. 
By doing this, we can guarantee the monotonicity property of the two-stage auction, that is when the advertiser $i$ increases her bid, she has a high probability to win in the pre-auction and enter the second stage, in which she obtains a not worse ad slot. 
Secondly, we formulate the optimization problem in the pre-auction stage. The goal of the pre-auction is to nominate a good candidate set of ads $\setA_M$ for the second stage GSP auction such that the expected social welfare is maximized. 
With the candidate ads set $\setA_M$, the GSP auction displays the top $K$ ads, denoted as the set $\setA_M^K$, with the highest expected click value, $b_i\times ctr(a_i,u)$, resulting in the expected social welfare $\stk( \{ b_i\times ctr_i \}_{i\in\setA_M} )=\sum_{i\in\setA_M^K} b_i \times ctr_i$. 
In the first stage, we have no access to the refined $ctr(a_i,u)$ but only the $(\tilde{a}_i,\tilde{u})$, and thus the optimization problem for the pre-auction is a stochastic optimization problem: 
\begin{equation}
  \label{eq:problem-pa}
\begin{array}{l}
\text{(PA)} \,\,
\displaystyle \max_{\bs{x}^p} \,
\E_{\bs{a}|\bs{ \tilde{a}},{u}|\tilde{u}}
[\stk( \{ b_i \times ctr_i(a_i, u) \}_{i\in \setA_M} )] \\
\quad\quad\quad\,
s.t. 
\,\, 
\setA_M=\bs{x}^p(\bs{b},\bs{\tilde{a}}, \tilde{u}), \\
\quad\quad\quad\quad\,\,\,
x^p_i(\bs{b},\bs{\tilde{a}},\tilde{u}) \text{ is monotone on } b_i , \forall \bs{b}_{-i}, \bs{\tilde{a}}, \tilde{u},
\end{array}
\end{equation}
where $\bs{b}_{-i}$ is the vector of bids after removing $b_i$ from $\bs{b}=(b_i,\bs{b}_{-i})$.

\section{Suboptimality of Greedy Solution}
\label{sec:analysis}

In this section, we show the performance degradation of a native greedy two-stage ad auction (GDY), which is widely used in industry. 
The GDY auction has a simple and intuitive definition based on the refined and coarse ctr estimators $\M^r={ctr}(a_i,u)$ and $\M^c=\widetilde{ctr}(\tilde{a}_i,\tilde{u})$.  
\begin{definition}
\label{def:greedy}
In GDY, the pre-auction stage ranks all the ads $\setA_N$ by $b_i \times \widetilde{ctr}(\tilde{a}_i,\tilde{u})$, and delivers the highest $M$ ads $\setA^g_M$ to the second stage, which runs a GSP auction on set $\setA^g_M$. 
\end{definition}
We investigate whether GDY is a proper solution for the two-stage ad auction. 
Only when $M=K$, the pre-auction in GDY is exactly the optimal solution for the problem in (\ref{eq:problem-pa}): 
\begin{equation*}
\label{eq:m=k}
\begin{array}{ll}
& \E [ \stk( \{b_i\times ctr_i\}_{i \in \setA_M} ) ]
= \E [ \sum_{i\in\setA_M} b_i\times ctr_i ] \\
= & \sum_{i\in\setA_M} \E [b_i\times ctr_i ]
= \sum_{i\in\setA_M} b_i\times \widetilde{ctr}_i,
\end{array}
\end{equation*}
where the first equality is due to $M=K$, the second equality is due to linearity of expectation, and the third equality comes from the relation between $ctr$ from $\M^r$ and $\widetilde{ctr}$ from $\M^c$ shown in (\ref{eq:ctr-relation}). 
Thus, selecting the ads with the largest $M$ values of $b_i\times \widetilde{ctr}_i$ to form $\setA_M$ in GDY is optimal in this case. 
For the case of $M>K$, we use a simple example to explain the suboptimality of GDY. 
\begin{example}
\label{eg:sub-optimal-greedy}
There are $N$ ads with 
$b_1\times \widetilde{ctr}_1 > \ldots > b_N \times \widetilde{ctr}_N$. 
$\forall i\leq M$, $ctr_i=\widetilde{ctr}_i$. 
An ad $j>M$ has two possible CTR realizations,  $ctr_j=t\times \widetilde{ctr}_j$ with probability $1/t$, and $ctr_j=\epsilon$ with probability $1-1/t$, where $t>0$ is a large number such that $b_j\times t\times \widetilde{ctr}_j > b_K\times \widetilde{ctr}_K = b_K\times ctr_K$.  
GDY first selects the top $M$ ads (without the ad $j$) in the pre-auction stage, and then displays the top $K$ ads to the user in the second stage. 
The resulted expected social welfare is $\sum_{i=1}^K b_i\times ctr_i$. 
However, if the pre-auction stage selects $\setA_M=\{ 1,\ldots,M-1,j \}$, then after the second stage GSP auction, the expected social welfare outperforms that of GDY: 
\begin{equation*}
  \label{eq:eg-better-set}
\begin{array}{ll}
& \E[ \stk( \{ b_i\times ctr_i \}_{i\in \setA_M}) ] \\
= & \frac{1}{t} ( \sum_{i=1}^{K-1} b_i\times ctr_i + b_j\times t \times \widetilde{ctr}_j ) 
+ (1-\frac{1}{t}) \sum_{i=1}^K b_i\times ctr_i \\
> & \sum_{i=1}^K b_i \times ctr_i. 
\end{array}
\end{equation*}
\end{example}


The greedy two-stage auction would encounter the scenarios similar to Example \ref{eg:sub-optimal-greedy}, resulting in performance degradation in practice. 
Some ad (like ad $j$ in the example) get a low coarse $\widetilde{ctr}$ from $\M^c$ but its refined $ctr$ from $\M^r$ is high.
If the pre-auction stage is aware that the second stage GSP auction will further refine the $ctr$ estimation, the pre-auction stage can make a better decision, selecting some ads with potential high refined ctr to achieve a higher social welfare.
This example uses the relation between the coarse estimator $\M^c$ and refined estimator $\M^r$ shown in (\ref{eq:ctr-relation}), \ie, the pre-auction stage can regard the unrealized refined $ctr(a_i,u)$ as a random variable from a distribution with the mean value $\widetilde{ctr}(\tilde{a}_i,\tilde{u})$. 
This problem is a subset selection over a set of random variables in the literature, and the theoretical analysis have been considered in team selection problem \cite{kleinberg2018team} and other general background \cite{mehta2020hitting}. 


The main insight we want to deliver here is that the widely deployed greedy solution, regarding the design in each of the two stages as  social welfare maximization separately and using the same selection metric ($b\times \widetilde{ctr}$ or $b\times ctr$) in both stages, would instead suffer a suboptimal social welfare. 
When design a two-stage auction, we should consider the interaction between the two stages, \ie, the second stage will refine the estimation and conduct the ad allocation over the subset delivered by the first stage, and design proper selection metrics for each stage, to guarantee the overall ad performance of the two-stage ad auction. 


\input{./21fupa_main_wsdm/21fupa_solution.tex}

\input{./21fupa_main_wsdm/21fupa_experiment.tex}


\input{./21fupa_main_wsdm/21fupa_related.tex}


\input{./21fupa_main_wsdm/21fupa_conclusion.tex}

%% file: 21fupa_main_wsdm/21fupa_fig_problem.tex
\begin{figure*}[!t]
\captionsetup[subfigure]{justification=centering}
\centering
\subfloat[Greedy design]{
\label{fig:problem-gdy}
\includegraphics[width=0.40\textwidth]{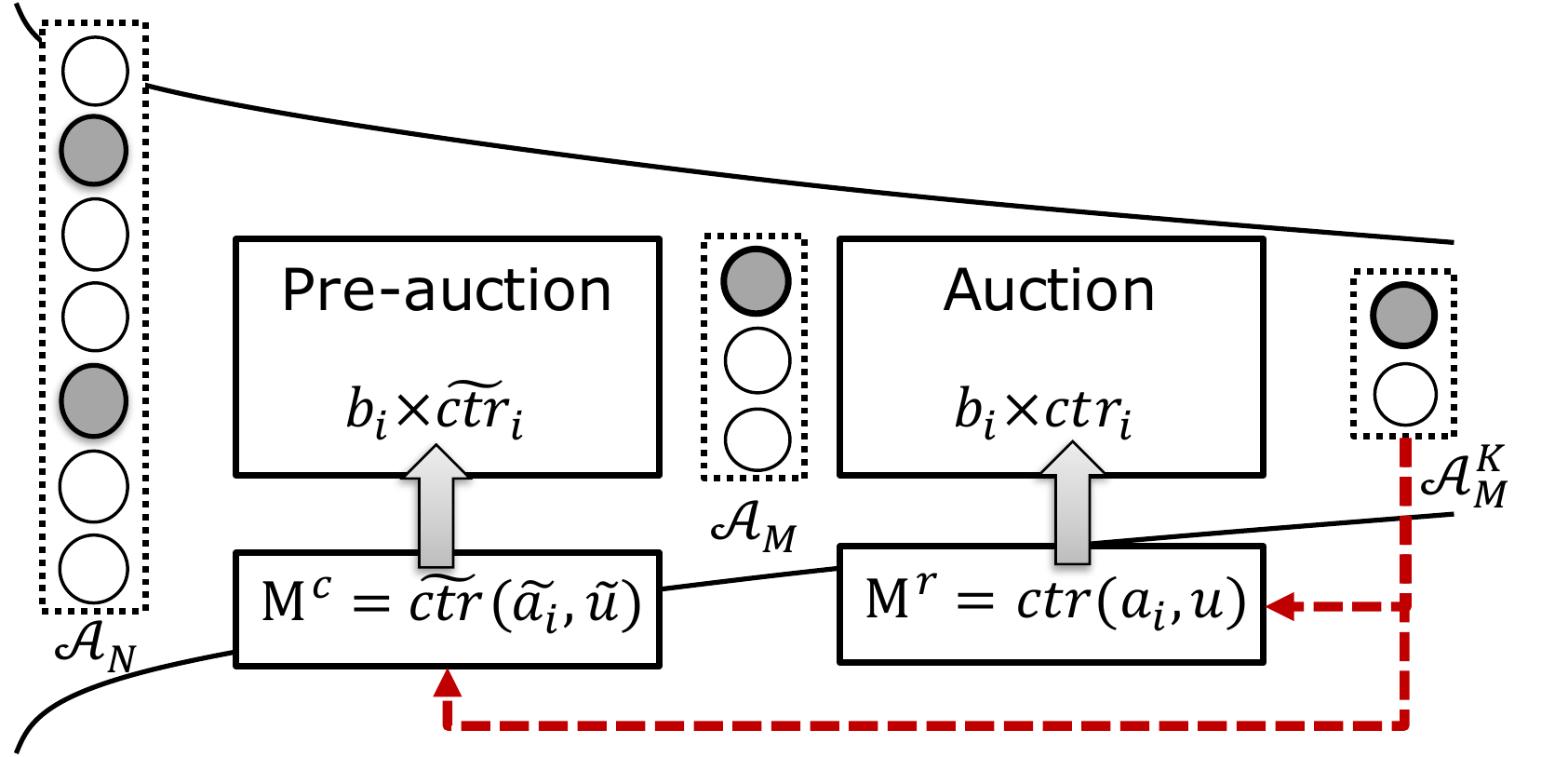}
}
\subfloat[PAS design]{
\label{fig:problem-pas}
\includegraphics[width=0.40\textwidth]{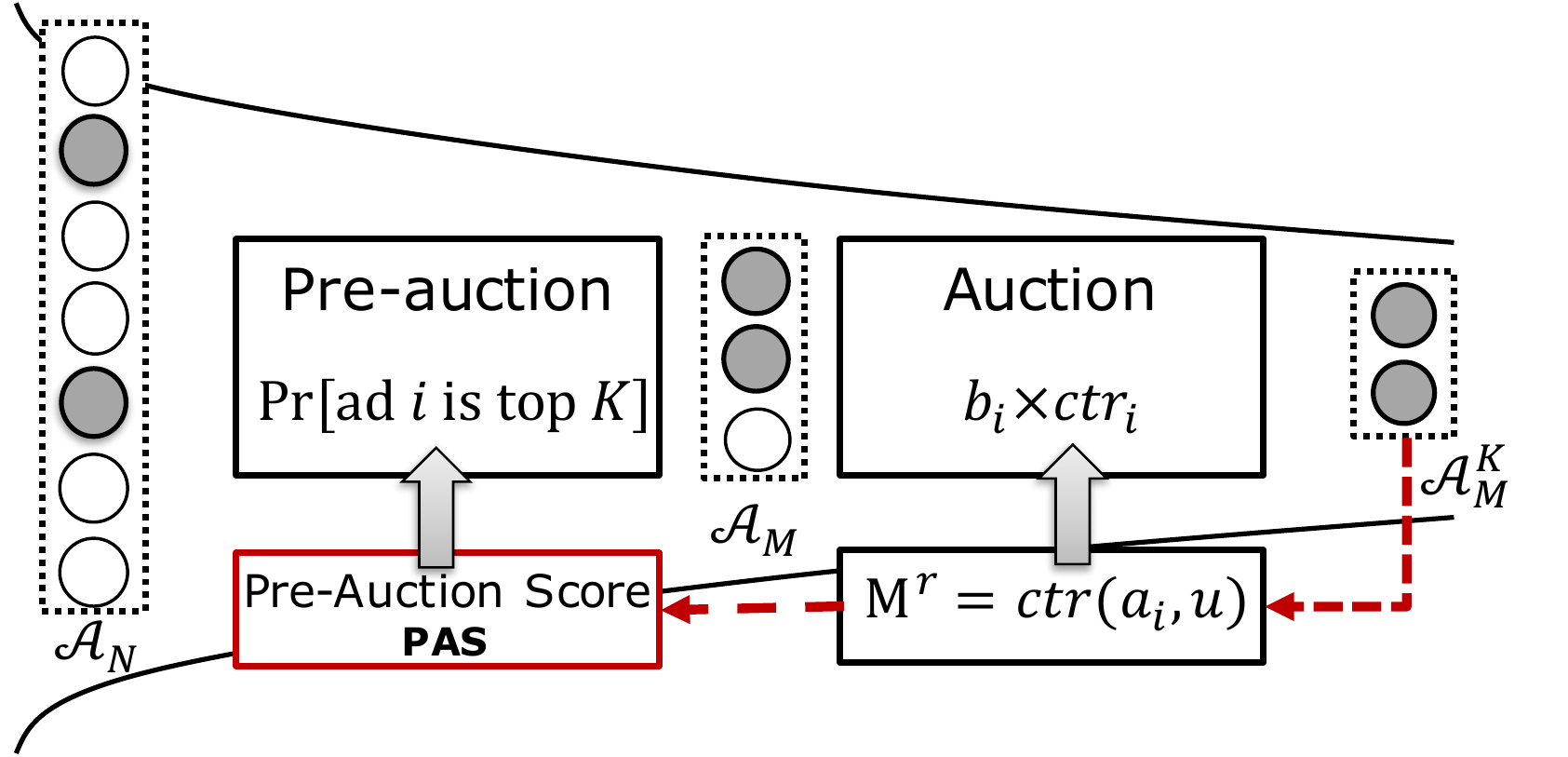}
}
\caption{
(a) The widely adopted greedy design of two-stage auction. (b) Our design with Pre-Auction Score (PAS). 
}
\label{fig:problem}
\end{figure*}

%% file: 21fupa_main_wsdm/21fupa_solution.tex
\section{Pre-auction Design}

We first show the computational complexity of solving the pre-auction problem defined in (\ref{eq:problem-pa}). 
To reduce the complexity, we then propose an ad-wise metric called pre-auction score (PAS) for scalable ad selection in the pre-auction stage. 
We also design a learning based implementation for PAS in practice. 
The detailed proofs in this section can be found in Appendix~\ref{appendix:proof}. 



\subsection{Complexity of Pre-auction Problem}


We first prove that, even we relax the constraint of monotonicity on the allocation $x_i^p$ of the problem PA in (\ref{eq:problem-pa}), the resulting simplified version of PA is still intractable. 
\begin{equation*}
\begin{array}{lll}
\text{(SimPA)} & \displaystyle \max_{\setA_M \subseteq \setA_N } & \E_{ \bs{ctr} }[ \stk( \{ b_i\times ctr_i \}_{ i\in \setA_M } ) ] \\
& s.t. &  | \setA_M | \leq M,
\end{array}  
\end{equation*}
where $ctr_i$ is a  random variable. 
\begin{proposition}
\label{prop:simpa-np-hard}
The SimPA problem is NP-hard. 
\end{proposition}
In fact, SimPA is a submodular maximization with a cardinality constraint. 
For a ground set $\setA$, a set function $g: 2^\setA \rightarrow \mathbb{R}$ is submodular if $\forall \mathcal{S} \subseteq \mathcal{T} \subset \setA$ and $\forall j\in \setA\backslash \mathcal{T}$, $g(\mathcal{S}\cup\{j\})-g(\mathcal{S})\geq g(\mathcal{T}\cup\{j\})-g(\mathcal{T})$.
\begin{proposition}
\label{prop:simpa-submodular}
In SimPA, the objective $\E_{\bs{ctr}} [ \stk ( \{ b_i\times ctr_i \}_{ i\in\mathcal{S} } ) ]$ is a submodular set function with respective to the set $\mathcal{S}$. 
\end{proposition}


There is no apparent tractable and scalable solution for this submodular optimization in the setting of pre-auction stage: 
(i) The brute force algorithm, which evaluates the selection $\setA_M$ at set-wise scale, causes an intractable $O({N \choose M})$ computation. 
(ii) The classical approximation algorithms~\cite{nemhauser1978analysis,buchbinder2014submodular} select ads sequentially based on their marginal contributions, failing to run in an ad-wise parallel way, and thus are still not scalable in practice. 
(iii) We emphasize that from the perspective of the pre-auction stage, $\bs{ctr}$ as well as its explicit distribution are unknown, which introduces difficulty on even evaluating the objective submodular function. 
Thus, even the parallel approximation algorithms~\cite{chekuri2019submodular}, which rely on submodular function evaluation, are not suitable here. 

Considering the scalability in online service of pre-auction stage, we propose an ad-wise metric for subset selection in Section \ref{sec:pas}. 
With this ad-wise metric, we can evaluate each ad in a parallel way, and avoid evaluation of the submodular set function. 
To tackle the lack of explicit distribution of $\bs{ctr}$, we propose learning based implementation for the ad-wise metric in Section \ref{sec:learning-based-pas}.  

\subsection{Pre-auction Score}
\label{sec:pas}


We design a tractable and scalable ad-wise metric for subset selection in the pre-auction stage, which supports parallel evaluation on each ad. 
Since that social welfare maximization of pre-auction is an NP-hard problem, it is hopeless to directly obtain an ad-wise metric as the exact solution for social welfare maximization. 
The key insight here is that we can turn to an objective closely related to social welfare maximization, and then derive a corresponding ad-wise metric. 
The closely related objective, called \emph{recall maximization} (PA-R), is to select the set of ads $\setA_M$ in the pre-auction stage to cover as much final top K ads of $\setA_N$ in the auction stage, 
\begin{equation}
  \label{eq:problem-par}
\begin{array}{l}
\text{(PA-R)} \,\, 
\displaystyle \max_{\setA_M \subseteq \setA_N} \,
\E_{\bs{ctr}}[\stk( \{ \mathbbm{1}[i\in\setA_N^K] \}_{i\in \setA_M} )], 
\end{array}
\end{equation}
where the final top $K$ ads of $\setA_N$, denoted as $\setA_N^K$, are the ads with the highest $K$ $b_i\times ctr(a_i,u)$ from $\setA_N$ in the auction stage. 
If $\setA_M$ covers all the top $K$ ads $\setA_N^K$, it achieves the maximum social welfare of the one-stage auction, \ie, the upper bound social welfare of two-stage ad auction. 


We now derive the exact ad-wise metric, such that the pre-auction can rank and retrieve the top $M$ ads $\setA_M$ according to this metric to optimize PA-R. 
Consider any fixed subset $\setA_M$, the expected recall on $\setA_N^K$ is 
\begin{equation}
\label{eq:derive-pas}
\begin{array}{ll}
& \E_{\bs{ctr}} 
[ \stk( \{ \mathbbm{1}[i\in\setA_N^K] \}_{i\in\setA_M}) ] \\
= & \sum_{i\in \setA_M} \E_{\bs{ctr}} 
[ \mathbbm{1}[i\in\setA_N^K] ] \\
= & \sum_{i\in \setA_M} \Pr_{\bs{ctr}}[i\in \setA_N^K],
\end{array}
\end{equation}
where the first equality is due to 
the linearity of expectation and the second equality is due to the definition of indicator function. 
The above equation shows that selecting the ads with the largest $M$ $\Pr_{\bs{ctr}}[i\in \setA_N^K]$, \ie, the probabilities of being in $\setA_N^K$, maximizes the expected recall on $\setA_N^K$. 
Hence, we obtain an ad-wise metric for subset selection in the pre-auction, denoted as $f_i$ for each ad $i$:
\begin{equation}
\label{eq:result-pas}
\begin{array}{l}
\text{(PAS)} \quad\quad\quad\quad f_i(\bs{b},\bs{\tilde{a}},\tilde{u}) = \Pr_{\bs{ctr}}[i\in \setA_N^K]. \quad\quad\quad
\end{array}
\end{equation}
We call this metric as \emph{Pre-Auction Score (PAS)}. 
Note that although PAS is an ad-wise metric, it is still allowed to use the information of the whole ad set $\setA_N$ in the pre-auction stage, \ie, $\bs{b}, \bs{\tilde{a}}, \tilde{u}$, to facilitate the calculation of the probability. 
According to definition of set $\setA_N^K$ and PAS in (\ref{eq:result-pas}), for any advertiser $i$, partial ads features $\bs{\tilde{a}}$, user features $\tilde{u}$ and other advertiser's bid $b_{-i}$, the advertiser $i$'s PAS is monotonely increasing with respective to $b_i$. 
Thus, if the pre-auction ranks all the ads $\setA_N$ with the metric PAS and selects the top $M$ ads, it satisfies the monotonicity of allocation in the pre-auction, and hence satisfies the IC of the two-stage auction. 


\subsection{Learning Based Pre-auction Score}
\label{sec:learning-based-pas}



Computing the metric PAS requires an explicit form of distribution $\Pr[\bs{ctr}|\bs{b},\bs{\tilde{a}},\tilde{u}]$, which might be complicated within the unknown and interdependent online environment. 
To overcome this difficulty, we use parametrized neural networks to implement a learning based PAS $f_i^\theta$, such that the permutation of ranking by $f_i^\theta$ can approximate the permutation of ranking by the original PAS $f_i$ in (\ref{eq:result-pas}). 
We use supervised learning to determine the parameters $\theta$ of the neural networks $f_i^\theta$. 
For each training sample, the features are $(\bs{b},\bs{\tilde{a}},\tilde{u})$, \ie, the $N$ bids, the partial ad features and the partial user features; 
and the label is the $N$-dim vector $\bs{y}$, where the element is $y_i=b_i\times ctr(a_i,u)$ for each ad $i$. 
During the training process, we need $ctr(a_i,u)$ from the refined estimator $\M^r$ for all the ads in $\setA_N$ to compute the label of a sample. But we only obtain $ctr(a_i,u)$ for ad $i\in \setA_M$ during the online service, because only the ads $\setA_M$ enter the second auction stage and are evaluated by the refined estimator.
Thus, we need an offline refined estimator $\M^r$ to produce $ctr(a_i,u)$ for the ads in $\setA_N\backslash\setA_M$. 


Follow the Plackett-Luce probability model, which is widely used in learning the distribution of permutations~\cite{cao2007learning, guiver2009bayesian}, we assume that the permutation $\pi$ of ranking by $y_i=b_i\times ctr(a_i,u)$ is a sample from the distribution of permutations which is defined with $N$ parameters $\{y_i\}_{i\in\setA}$ as follows
\begin{equation}
\label{eq:permutation-prob}
\Pr [\pi | \bs{y}] = \prod_{i=1}^N \frac{y_{\pi(i)}}{\sum_{k=i}^N y_{\pi(k)}},
\end{equation}
where $\pi(k)$ is the ad with rank $k$. 
We can easily verified that this definition of probability satisfies two desired properties: 
(i) It is normalized, \ie, $\sum_{\pi} \Pr[\pi | \bs{y}] = 1$; 
(ii) The top 1 probability is: 
\begin{equation}
\label{eq:top-1-probability}
\Pr[i \text{ is the top 1 } | \bs{y}] = \frac{y_i}{\sum_{k=1}^N y_k}.
\end{equation}
\begin{lemma}
\label{lem:relation-y-topk}
For $y_i>y_j$ and $K$ > 1, $\Pr[i \in \setA_N^K] > \Pr[j \in \setA_N^K]$.
\end{lemma}
With Lemma.\ref{lem:relation-y-topk}, we can use the rank by $y_i$ to represent the rank by PAS $\Pr[i \in \setA_N^K]$. 
Let the neural networks directly output $f_i^\theta$ as logits for approximate $y_i$. 
The top-1 probability of the ad $i$ determined by logits $\{f_i^\theta\}_{i\in\setA_N}$ is calculated as
$$
\Pr[i \text{ is the top 1 } | \bs{f}^\theta(\bs{b}, \bs{\tilde{a}}, \tilde{u})] = \frac{\exp(f_i^\theta)}{\sum_{k\in\setA_N} \exp(f_k^\theta)}.
$$
Given the sample set $\mathcal{D}_f$, we minimize the cross entropy between the sample distribution and the parametric distribution by $\bs{f}^\theta$, so the loss function is
\begin{equation}
\label{eq:loss-softmax}
\begin{aligned}
L = 
-\frac{1}{|\mathcal{D}_f|} \sum_{j\in\mathcal{D}_f} \sum_{i=1}^N 
&  
\Pr[i \text{ is the top 1} | \bs{y}^j]  \\
& \times \log \Pr[ i \text{ is the top 1} | \bs{f}^\theta(\bs{b}^j, \bs{\tilde{a}}^j, \tilde{u}^j) ] 
\end{aligned}
,
\end{equation}
where the superscript $j$ means the $j$-th sample in $\mathcal{D}_f$. 


%% file: 21fupa_main_wsdm/21fupa_experiment.tex
\section{Experiments}

We provide empirical evidence for the effectiveness of our proposed two-stage auction solution on both public and industrial datasets. 

\subsection{Settings for Public Dataset}


The public dataset Amazon Dataset \footnote{http://jmcauley.ucsd.edu/data/amazon/} contains product reviews and metadata from Amazon \cite{he2016ups,mcauley2015image}. 
We conduct experiments on the subset called Books, which contains 603K user reviews, 367K items and 1600 categories.
We regard reviews as user clicks and regard items as ads.  
The full features of a user $u$ include a user\_id, and a list of user's reviewed items along with categories in history, \ie, $u=\langle \text{user\_id, hist\_items\_id, hist\_cate\_id} \rangle$. 
The length of each user's item list is at least 5. 
The full ad features $a_i$ is $a_i=\langle \text{item\_id, cate\_id} \rangle$. 
We define the partial user feature $\tilde{u}$ as $\tilde{u}=\langle \text{user\_id, hist\_items\_id[:3], hist\_cate\_id[:3]} \rangle$, \ie, only the first three items in user's history are used in $\tilde{u}$. 
The partial ad features $\tilde{a}_i$ is defined as the same as the full ad features $a_i$. 
Based on this dataset, we simulate the process of two-stage auction and generate 40,000 auctions. 
In each auction, there are 1000 randomly selected ads. 
The bid of each ad is independently sampled from uniform distribution.


To simulate the second stage GSP auction, we use ad and user's full features $\langle a_i,u \rangle$ to train Deep Interested Networks (DIN), a baseline $ctr$ estimator \cite{zhou2018deep}, as the refined estimator $\M^r$ to generate $ctr(a_i,u)$, and the label $b_i\times ctr(a_i,u)$ for each ad. 
To simulate GDY for comparison, we use ad and user's partial features $\langle \tilde{a}_i,\tilde{u} \rangle$ to train a model with embedding followed by fully connected layers (FCN) \cite{wang2020cold}, as the coarse and fast estimator $\M^c$ to generate $\widetilde{ctr}(\tilde{a}_i,\tilde{u})$ for each ad.  
The training samples for $\M^r$ DIN and $\M^c$ FCN are the same, \ie, the same $\langle \text{ad, user} \rangle$ pairs from the dataset as positive samples, along with the same 1:1 negative sampling. 
Let the negative down sampling rate be $\eta$. 
When a estimator trained with 1:1 negative samples outputs $p$, then the resulting $ctr=\frac{p}{p+(1-p)/\eta}$ \cite{he2014practical}. 
The difference between $ctr$ and $\widetilde{ctr}$ gets larger as $\eta$ decreases, because the difference between raw output $p$ from $\M^r$ and $\M^c$ is approximately amplified by the factor $1/\eta$. 
We consider two settings, Public-1 and Public-5, where the negative down sampling rates are $0.01$ and $0.05$, respectively. 
Public-1 simulates the $ctr$ in real-world online advertising. 
We would like to verify whether our solution can still outperform GDY in the scenario simulated by Public-5, where the difference between $ctr$ and $\widetilde{ctr}$ is small and GDY are able to achieve a relatively better performance.

\subsection{Settings for Industrial Dataset}



The industrial dataset comes from the log of a two-stage ad auction in a leading e-commerce platform, running GDY defined as Definition \ref{def:greedy}. 
We randomly sample 30K auction records from the logged data on January 8th 2021. 
There are 700 ads in each auction instance. 
For each ad, the features are: 
(i) $\widetilde{ctr}(\tilde{a}_i,\tilde{u})$ from $\M^c$, some historical statistics such as historical averaged refined $ctr$ and $cvr$ of this ad. 
We regard these estimated values as cross features of the ad and user's partial features $\langle \tilde{a}_i,\tilde{u} \rangle$; 
(ii) Ad information like bid $b_i$, category and selling price of the product in the ad. 
The estimated $ctr(a_i,u)$ from $\M^r$ are used to generate the label $b_i\times ctr(a_i,u)$. 
Note that under our assumption of value maximizing advertisers, GDY satisfies IC and IR. 
Therefore, we can regard the logged bids as advertiser's truthful values for user clicks.
Then, we can use the logged bids to compute social welfare and revenue during simulating other IC and IR auction mechanisms. 

\input{./21fupa_main_wsdm/21fupa_tab_perf_pub_01.tex}
\input{./21fupa_main_wsdm/21fupa_tab_perf_pub_05.tex}
\input{./21fupa_main_wsdm/21fupa_tab_perf_ind.tex}

\subsection{Evaluation Metrics}


We consider the following common used metrics for ad auction evaluation. 
We recall that $\setA_N^K$ is the true top $K$ ads in the whole set $\setA_N$ while $\setA_M^K$ is the top $K$ ads in the selected subset $\setA_M$. 
\begin{itemize}
\item Social welfare rate: $SWr@K=\frac{\sum_{i\in \setA_M^K}b_i \times ctr_i}{\sum_{j\in \setA_N^K} b_j \times ctr_j}$.
\item Top K Recall: $Recall@K={\sum_{i\in \setA_M^K}\mathbf{1}[i\in \setA_N^K]} / K$. 
\item Revenue rate: $REVr@K=REV(\setA_M^K)/REV(\setA_N^K)$, where revenue $REV$ is under GSP auction and the subscript $(k)$ means the advertiser with the $k$-th highest $b \times ctr$ in the corresponding set, 
and 
$REV(\setA)=\sum_{k=1}^K b_{(k+1)}\times ctr_{(k+1)}.$
\end{itemize}
\subsection{Methods for Comparison}



To prove the effectiveness of our proposed solution with PAS metric and its learning based implementation, we introduce the following baselines of pre-auction for comparison. 
Due to the practical deployment requirement, we only focus on the methods work as \emph{ranking by an ad-wise metric} to select a subset of $M$ ads from the $\setA_N$, and we only describe their selection metrics here. 
Implementation of these ad-wise metrics are restricted to use the same partial features $\langle \tilde{a},u \rangle$ as PAS uses. 
\begin{itemize}
\item Greedy (GDY): As described in Definition \ref{def:greedy}, the rank score of GDY for ad $i$ is simply $b_i\times \widetilde{ctr}_i$, where $\widetilde{ctr}_i$ is the output by the coarse estimator $\M^c$. 
\item Regression to $ctr_i$ (REGCTR): We use $ctr_i$ as label, and use ad and user's partial features $\langle \tilde{a}_i$, $\tilde{u} \rangle$ to train a regression model with mean square loss. 
The rank score is bid times output of the regression model.
\item Regression to $b_i \times ctr_i$ (REG): We use $b_i\times ctr_i$ as label, and use the partial ad, user features and the bid, \ie, $b_i,\tilde{a}_i, \tilde{u}$, to train a regression model with mean square loss. 
\end{itemize}
The neural network structure and input for REGCTR, REG, and PAS are almost the same, except that REGCTR network lacks the input of bid. 
The reason we introduce REGCTR and REG for comparison is to demonstrate that PAS is a more proper selection metric for pre-auction. 
In each repeated experiment, we split data into training, validation, and test set with 3:1:1, and apply early stopping with metric $SWr@5$ on validation set for all three methods. 

\subsection{Performance Comparison}


Results of different methods on data setting Public-1, Public-5 and Industrial are given in 
Table \ref{tab:perf-pub-01}-\ref{tab:perf-ind}. 
The tables show average metrics, standard deviation and the improvement over GDY in 20 runs with 20 distinctive random seeds for each data setting and methods. 


We can obtain the following observations from Table \ref{tab:perf-pub-01}-\ref{tab:perf-ind}. 
(i) We can see that our proposed PAS outperforms all baseline methods in each data setting and on each metric. 
For instance, PAS improves $SWr@5$ by $+3.61\%$, $+3.31\%$ and $+4.35\%$ comparing with the widely used GDY in Public-1, Public-5 and Industrial, respectively. 
(ii) As REGCTR, REG and PAS have better performance than GDY on most data settings and metrics, we conclude that pre-auction stage can benefit from the supervision by the second stage's information. 
(iii) REGCTR and REG fall behind our proposed PAS. 
The reason can be that while REGCTR and REG are forced to learn regression on $ctr_i$ or $b_i\times ctr_i$, PAS who models the probability of being top $K$, is a more proper selection metric for the pre-auction stage. 


Next, we compare the performance results in Table \ref{tab:perf-pub-01}-\ref{tab:perf-pub-05} for Public-1 and Public-5, who only differ on their down negative sampling rates $\eta$. 
The smaller the $\eta$ is, the larger the gap between coarse $\widetilde{ctr}$ and refined $ctr$ is. 
This indicates that GDY on Public-1 has a worse performance than on Public-5, which can also be observed from the the evaluation results. 
We can also see that REGCTR and REG achieve much worse performance on Public-1 than on Public-5. 
For example, $SWr@5$ of REG on Public-5 is $0.967$ which outperforms $SWr@5$ of GDY by $+2.84\%$, 
but the corresponding values turn out to be only $0.927$ and $+0.22\%$ on Public-1. 
However, PAS are more stable than REG and REGCTR, and achieves similar good results on both Public-1 and Public-5.




\input{./21fupa_main_wsdm/21fupa_fig_br.tex}

In Figure \ref{fig:session-plot}, we plot an auction instance for each of the three data settings to show the results of the four methods work in a two-stage auction. 
The figures for Public-1 and Public-5 are from the same auction instance, while only the down negative sampling rates for calculating $ctr$ are different. 
We first explain the three large figures in the above of Figure \ref{fig:session-plot} which illustrate the results of method GDY. 
The x-axis is the rank by $bid\times ctr$ with the refined $ctr$ estimator, and here we show the top 200 ads. 
A blue and red point with the same x coordinate are associated with the same ad. 
The y-axis for blue points is the normalized value of $bid\times ctr$, while the y-axis for red points is the normalized value of $bid\times \widetilde{ctr}$. 
The black horizontal line shows the threshold of the normalized value $bid\times \widetilde{ctr}$ for entering the second auction stage in GDY. 
Red points above the black line enter the second auction stage, and the most left 5 red points among them obtain the top 5 ranks in the second auction stage and win ad slots. 
Next, we explain the small figures which show the top 20 ads in four different methods. 
Similarly, y-axis for the red points is the normalized score of pre-auction in the method, while the black horizontal line is the corresponding threshold for entering the second stage. 
We can see that variant methods result in different rankings in the pre-auction and different black horizontal lines. 
For example in the third column for Industrial data, the ad slots are allocated to the ads with the rank of $bid \times ctr$ as $\{2, 11, 15, 17, 20\}$ (GDY), $\{1, 2, 4, 11, 13\}$ (PAS), $\{1, 2, 8, 11, 13\}$ (REGCTR), and $\{1, 2, 8, 11, 13\}$ (REG), and thus PAS have a better $Recall@5$ and $SWr@5$ than GDY, REGCTR, and REG in this example. 

\subsection{IC Testing}

\input{./21fupa_main_wsdm/21fupa_tab_ic_test.tex}


As we have mentioned before that in order to guarantee the economic properties of IC and IR for the auction, we require that the allocation for each advertiser is monotonely increasing with respective to her bid. 
That is to say, the pre-auction stage need to satisfy: for any advertiser $i$, given partial features for all ads $\bs{\tilde{a}}$ and for the user $\tilde{u}$, as well as bids from other advertisers $b_{-i}$, there exists a threshold $b_i^t$ that the ad $i$ enters the second stage auction if and only if $b_i\geq b_i^t$. 
To show that learning based PAS can achieve the monotonicity approximately, we conduct counter factual perturbation on each advertiser's bid in the logged data and evaluate the violation of monotonicity condition. 
This is a common method for IC testing of ad auction mechanisms \cite{deng2019testing,deng2020data}. 


Specifically, we sample 1000 auctions from the test set. 
One IC test is defined on an auction and an ad.
For an ad $i$, all its features to PAS model remain the same except that we replace $b_i$ with $\alpha\times bid_i$, where $\alpha\in \mathcal{S}_p=\{0.2 \times j|j=1,\ldots,10\}$ is a multiplicative perturbation factor from interval $[0, 2]$. 
All the features of other ads in this auction remains the same. 
We simulate the two-stage auction on all 10 perturbation factors to check whether the PAS model pass the perturbation tests, \ie, (i) $\exists \underline{\alpha}\in \mathcal{S}_p$ such that ad $i$ can enter the auction stage with $\alpha\times b_i$ $\forall \alpha \geq \underline{\alpha}$; or (ii) ad $i$ can not enter auction stage with $\alpha\times b_i$ for any $\alpha\in \mathcal{S}_p$. 


We conduct $1000\times 1000$ tests, for 1000 auctions and 1000 advertisers per auction, in each of Public-1 and Public-5, and $1000\times 700$ in Industrial data. 
Table \ref{tab:ic-test} shows the average failure rates and the standard deviations of 20 runs. 
The low failure rates show that even we apply no deliberate design on PAS model for a guarantee of strict monotonicity, the learning based PAS learns an approximate monotonicity automatically. 
An intuitive reason might be that there is a signal in the supervised data that bid has a positive effect on the objective of model. 
We also test REG and obtain similarly low failure rates. 
For GDY and REGCTR, their metrics are monotone with respective to bid, so their allocations of pre-auction are naturally monotone.


%% file: 21fupa_main_wsdm/21fupa_tab_perf_pub_01.tex
\begin{table*}[!ht]
\caption{
Results of different methods on data setting Public-1. $N=1000, M=50$. Notations in a table cell: average $\pm$ standard deviation (Improvement over GDY) in 20 runs. 
}
\small
\centering
\begin{tabular}{|c|c|c|c|c|c|}
\hline
  &  $Recall@1$  &  $Recall@5$  &  $Recall@10$  &  $SWr@5$  &  $REVr@5$ 
\\ \hline
GDY	&
0.8383 $\pm$ 0.0022, \,\quad (0\%) &
0.7378 $\pm$ 0.0024, \,\quad (0\%) &
0.6597 $\pm$ 0.0029, \,\quad (0\%) &
0.9252 $\pm$ 0.0007, \,\quad (0\%) &
0.9028 $\pm$ 0.0010, \,\quad (0\%)
\\ \hline
REGCTR & 
0.8440 $\pm$ 0.0058, (+0.57\%) &
0.7469 $\pm$ 0.0079, (+0.91\%) &
0.6698 $\pm$ 0.0079, (+1.01\%) &
0.9234 $\pm$ 0.0048, (-0.18\%) &
0.9026 $\pm$ 0.0050, (-0.02\%)
\\ \hline
REG	& 
0.8506 $\pm$ 0.0068, (+1.23\%) &
0.7483 $\pm$ 0.0109, (+1.05\%) &
0.6709 $\pm$ 0.0112, (+1.12\%) &
0.9274 $\pm$ 0.0041, (+0.22\%) &
0.9058 $\pm$ 0.0050, (+0.30\%)
\\ \hline
\textbf{PAS} &
\textbf{0.9093 $\pm$ 0.0021, (+7.10\%)} &
\textbf{0.8639 $\pm$ 0.0045, (+12.61\%)} &
\textbf{0.8246 $\pm$ 0.0057, (+16.49\%)} &
\textbf{0.9613 $\pm$ 0.0011, (+3.61\%)} &
\textbf{0.9519 $\pm$ 0.0015, (+4.91\%)}
\\ \hline
\end{tabular}
\label{tab:perf-pub-01}
\end{table*}

%% file: 21fupa_main_wsdm/21fupa_tab_perf_pub_05.tex
\begin{table*}[!ht]
\caption{
Results of different methods on data setting Public-5. $N=1000, M=50$.
Notations are the same as Table.\ref{tab:perf-pub-01}.
}
\small
\centering
\begin{tabular}{|c|c|c|c|c|c|}
\hline
  &  $Recall@1$  &  $Recall@5$  &  $Recall@10$  &  $SWr@5$  &  $REVr@5$ 
\\ \hline
GDY	&
0.8359 $\pm$ 0.0023, \,\quad(0\%) &
0.7390 $\pm$ 0.0026, \,\quad(0\%) &
0.6623 $\pm$ 0.0025, \,\quad(0\%) &
0.9389 $\pm$ 0.0006, \,\quad(0\%) &
0.9213 $\pm$ 0.0008, \,\quad(0\%)
\\ \hline
REGCTR & 
0.9073 $\pm$ 0.0029, (+7.14\%) &
0.8502 $\pm$ 0.0028, (+11.12\%) &
0.7972 $\pm$ 0.0032, (+13.49\%) &
0.9669 $\pm$ 0.0008, (+2.80\%) &
0.9578 $\pm$ 0.0010, (+3.65\%) 
\\ \hline
REG	& 
0.9094 $\pm$ 0.0037, (+7.35\%) &
0.8521 $\pm$ 0.0051, (+11.31\%) &
0.7982 $\pm$ 0.0057, (+13.59\%) &
0.9673 $\pm$ 0.0016, (+2.84\%) &
0.9583 $\pm$ 0.0018, (+3.70\%)
\\ \hline
\textbf{PAS} &
\textbf{0.9155 $\pm$ 0.0071, (+7.96\%)} &
\textbf{0.8745 $\pm$ 0.0075, (+13.55\%)} &
\textbf{0.8363 $\pm$ 0.0080, (+17.40\%)} &
\textbf{0.9720 $\pm$ 0.0022, (+3.31\%)} &
\textbf{0.9651 $\pm$ 0.0027, (+4.38\%)}
\\ \hline
\end{tabular}
\label{tab:perf-pub-05}
\end{table*}

%% file: 21fupa_main_wsdm/21fupa_tab_perf_ind.tex
\begin{table*}[!ht]
\caption{
Results of different methods on data setting Industrial. $N=700, M=30$.
Notations are the same as Table.\ref{tab:perf-pub-01}.
}
\small
\centering
\begin{tabular}{|c|c|c|c|c|c|}
\hline
  & $Recall@1$ & $Recall@5$ & $Recall@10$ & $SWr@5$ & $REVr@5$ 
\\ \hline
GDY	&
0.4745 $\pm$ 0.0046, \,\quad (0\%) & 
0.3740 $\pm$ 0.0019, \,\quad (0\%) & 
0.3175 $\pm$ 0.0013, \,\quad (0\%) & 
0.7808 $\pm$ 0.0013, \,\quad (0\%) & 
0.7483 $\pm$ 0.0010, \,\quad (0\%) 
\\ \hline
REGCTR &
0.5156 $\pm$ 0.0047, (+4.11\%) &
0.4043 $\pm$ 0.0028, (+3.03\%) & 
0.3451 $\pm$ 0.0024, (+2.76\%) &
0.8081 $\pm$ 0.0017, (+2.73\%) &
0.7764 $\pm$ 0.0022, (+2.81\%)
\\ \hline
REG & 
0.5118 $\pm$ 0.0059, (+3.73\%) &
0.4048 $\pm$ 0.0033, (+3.08\%) &
0.3474 $\pm$ 0.0027, (+2.99\%) &
0.8089 $\pm$ 0.0022, (+2.81\%) &
0.7783 $\pm$ 0.0025, (+3.00\%)
\\ \hline
\textbf{PAS} &
\textbf{0.5351 $\pm$ 0.0035, (+6.06\%)} &
\textbf{0.4226 $\pm$ 0.0024, (+4.86\%)} &
\textbf{0.3635 $\pm$ 0.0017, (+4.60\%)} &
\textbf{0.8243 $\pm$ 0.0013, (+4.35\%)} &
\textbf{0.7942 $\pm$ 0.0007, (+4.59\%)}
\\ \hline

\end{tabular}
\label{tab:perf-ind}
\end{table*}

%% file: 21fupa_main_wsdm/21fupa_fig_br.tex
\begin{figure*}[!t]
\begin{minipage}{0.33\textwidth}
\includegraphics[width=\linewidth]{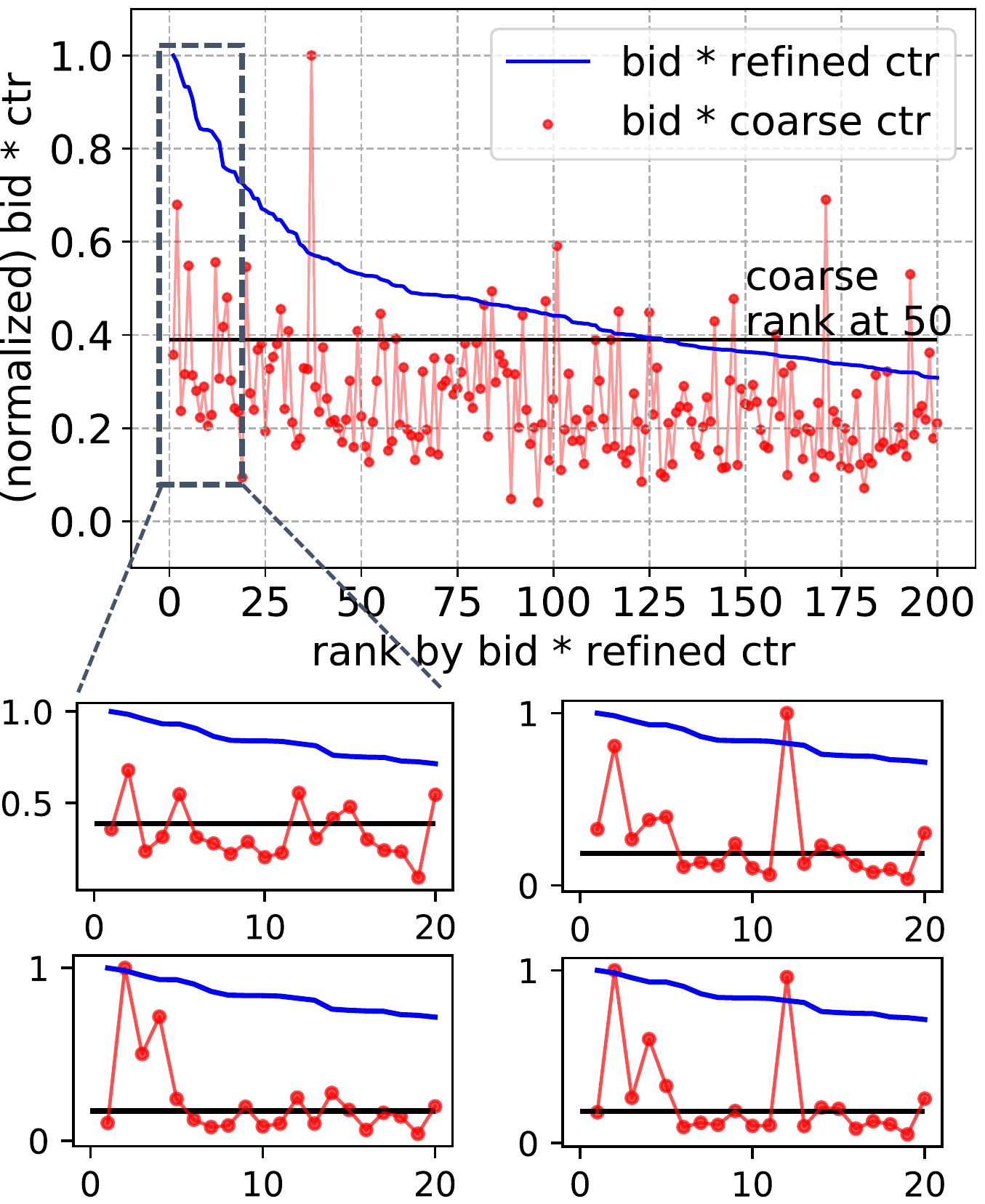}
\caption*{Public-1}
\label{fig:sess-public-1}
\end{minipage}\hfill
\begin{minipage}{0.33\textwidth}
\includegraphics[width=\linewidth]{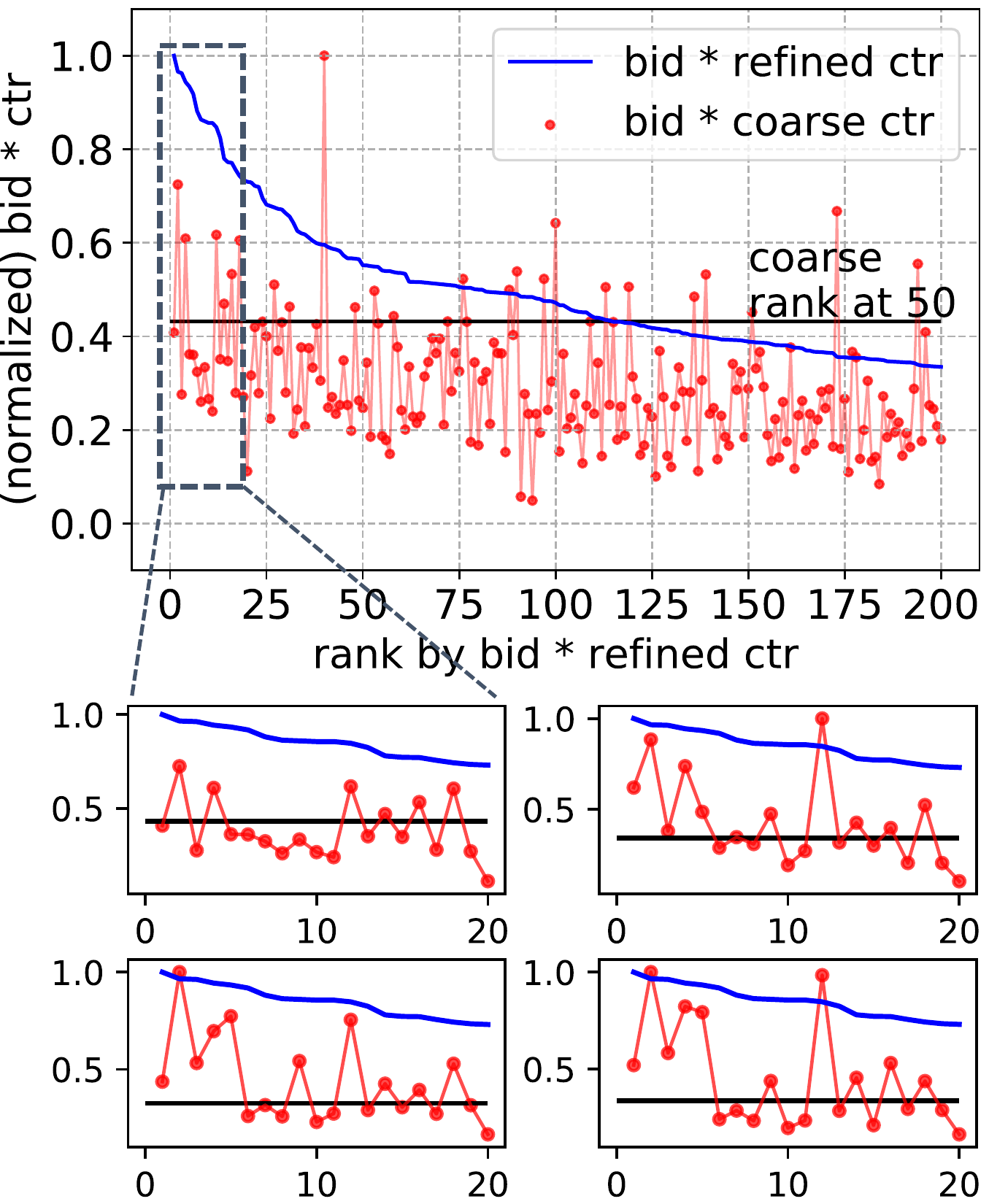}
\caption*{Public-5}
\label{fig:sess-public-5}
\end{minipage}\hfill
\begin{minipage}{0.33\textwidth}
\includegraphics[width=\linewidth]{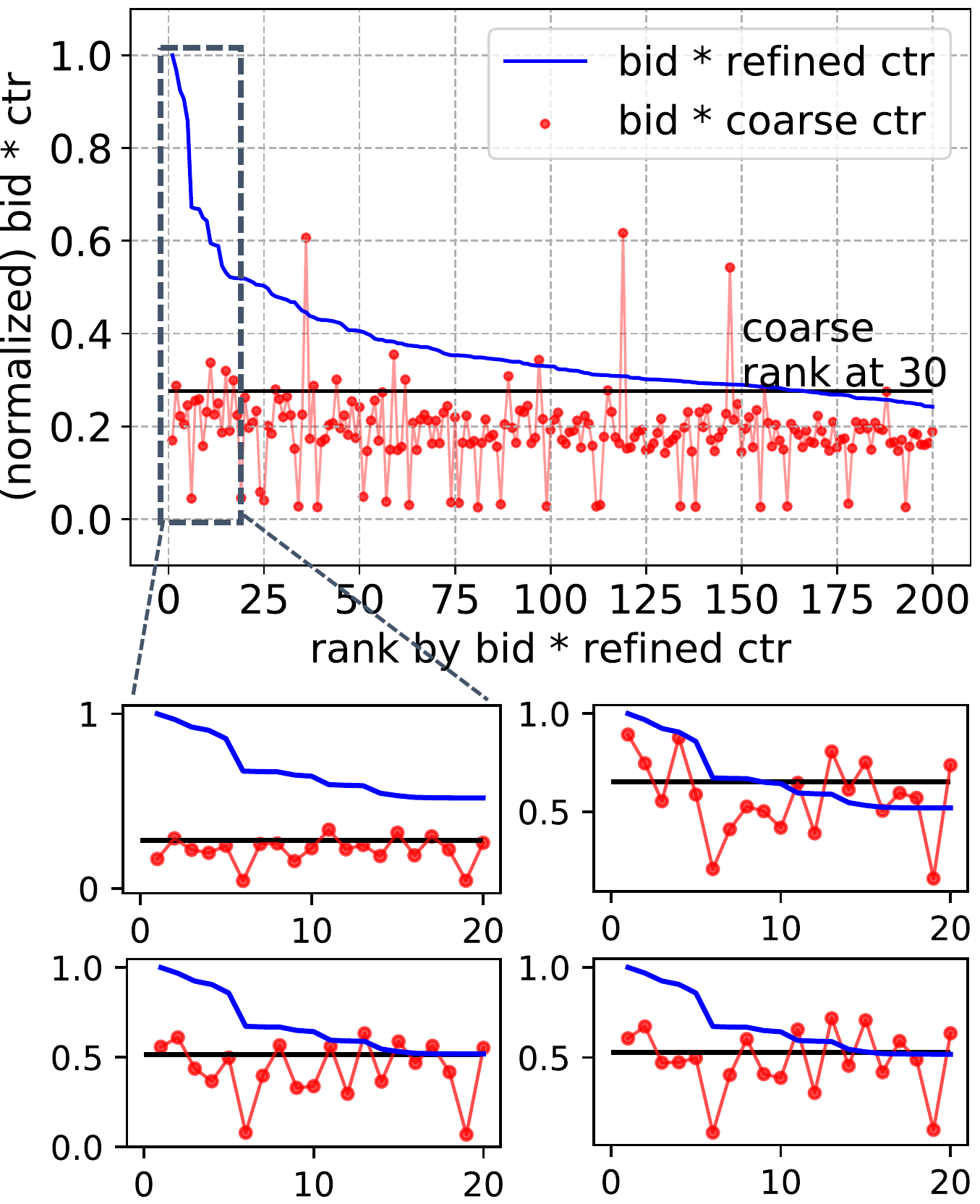}
\caption*{Industrial}
\label{fig:sess-industrial}
\end{minipage}
\caption{
Illustration of two-stage auction with methods GDY, PAS, REGCTR and REG on data Public-1, Public-5 and Industrial. 
Layout of the four small figures: top-left, GDY; top-right, PAS; bottom-left, REGCTR; bottom-right, REG. 
}
\label{fig:session-plot}
\end{figure*}

%% file: 21fupa_main_wsdm/21fupa_tab_ic_test.tex
\begin{table}[!t]
\caption{Failure rates of perturbation tests, with unit $\times 10^{-5}$. Average $\pm$ standard deviation in 20 runs.}
\small
\centering
\begin{tabular}{|c|c|c|c|c|}
\hline
        & GDY & REGCTR & REG & PAS
\\ \hline
Public-1 & 
0 &
0 & 
$10.4 \pm 5.27$ & 
$0 \pm 0$
\\ \hline
Public-5 & 
0 &
0 & 
$0 \pm 0$ & 
$0 \pm 0$
\\ \hline
Industrial & 
0 &
0 & 
$4.63 \pm 5.69$ & 
$1.40 \pm 2.34$
\\ \hline
\end{tabular}
\label{tab:ic-test}
\end{table}

%% file: 21fupa_main_wsdm/21fupa_related.tex
\section{Related Work}
\label{sec:related}

\textbf{Online advertising auction.} 
In the market of online ad auction, traditional auction mechanisms like general second price auction (GSP)~\cite{edelman2007internet,varian2007position} and Vickrey-Clark-Groove auction (VCG)~\cite{varian2014vcg} are widely used.
Recently, many parametric mechanisms learning from data are proposed to optimize performance metrics of the auction market, \eg squashed GSP~\cite{lahaie2007revenue}, boosted second price auction~\cite{golrezaei2017boosted}, and dynamic reserve price via reinforcement learning~\cite{shen2020reinforcement} for the task of revenue maximization. The paradigm of differentiable mechanism design via deep learning proposed by D{\"u}tting \etal~\cite{dutting2019optimal} is also applied to multi-objective optimization in online ad auctions~\cite{liu2021neural}.

\textbf{Two-stage recommender.} 
Two-stage architectures with candidate generation followed by ranking have been widely adopted in large scale industrial recommenders. 
Despite their popularity, the literature on two-stage recommenders is relatively scarce. 
To improve computation efficiency \cite{kang2019candidate,yi2019sampling} and recommendation quality \cite{chen2019top} are still the two most important problems of recommenders even in two-stage setting. 
Interaction between the two stages and its influence on system performance are studied in \cite{chen2019top,ma2020off,hron2021component}. 
Our problem of two-stage auction system differs from the two-stage recommenders. Since the payment transfer and requirement of economic properties, an auction system need quantifiable and explainable evaluation of user's interests like $ctr$, while a recommender only determines the order of items.

%% file: 21fupa_main_wsdm/21fupa_conclusion.tex
\section{Conclusion}


We have studied a novel problem of designing a large-scale two-stage ad auction, which consists of a pre-auction stage and an auction stage, to maximize social welfare for online advertising. 
We illustrated social welfare loss of a widely adopted design due to its improper selection metric in the pre-auction stage, which ignores the relation between the two auction stages. 
We have designed an IC and IR two-stage ad auction solution for value maximizers, where the second stage is GSP auction and the pre-auction stage selects the ad subset with an ad-wise metric called Pre-Auction Score (PAS).
We have further proposed a learning based implementation of PAS. 
Experiment results on both public and industrial dataset have shown that our proposed solution outperforms the greedy design significantly on social welfare and revenue of the two-stage ad auction.  

%% file: 21fupa_main_wsdm/21fupa_appendix.tex




\section{Proofs}

\label{appendix:proof}

\begin{proof}[Proof for Proposition \ref{prop:simpa-np-hard}]
We prove that SimPA is NP-hard even when $K=1$ by reducing a well-known NP-hard problem set cover to SimPA. 
We first define the general set cover problem. The universal set is $\mathcal{U}=\{u_1, \ldots, u_L\}$. There is a collection $\mathcal{T}=\{\mathcal{S}_1, \ldots, \mathcal{S}_N\}$ where $\forall i\leq N$, $\mathcal{S}_i\subset \mathcal{U}$ is a subset of universal set $\mathcal{U}$. 
The set cover problem is to decide whether there is a $\mathcal{T}'\subset \mathcal{T}, |\mathcal{T}'|\leq M$ such that 
$\bigcup_{\mathcal{S}_i\in \mathcal{T}'} \mathcal{S}_i = \mathcal{U}$. 
Next, for any instance of the set cover problem, we define the corresponding SimPA instance. 
Randomly sample a non-empty subset $\mathcal{U}'$ from universal set $\mathcal{U}$, then we define a set of random variables $\bs{ctr}=\{ctr_1, \ldots, ctr_N\}$ where each $ctr_i \in \{ 0, \frac{1}{b_i} \}$ is corresponding to the above set $\mathcal{S}_i$: if any element in $\mathcal{S}_i$ are sampled in $\mathcal{U}'$, $ctr_i=\frac{1}{b_i}$ and $b_i\times ctr_i=1$; otherwise, $ctr_i=0$ and $b_i\times ctr_i=0$. 
Therefore, there exists a $\mathcal{T}'\subset \mathcal{T}, |\mathcal{T}'|\leq M$ if and only if there exists an $\setA_M$ that $i\in \setA_M,\forall i\in \mathcal{T}'$ and $i\notin \setA_M$ otherwise, such that 
$\E_{ \bs{ctr} }[ \stk( \{ b_i \times ctr_i \}_{\setA_M} )]=1$ for $K=1$. 
\end{proof}
\begin{proof}[Proof for Proposition \ref{prop:simpa-submodular}]
We can easily verify that $\stk$ is a submodular set function. 
For any fixed value of $\bs{ctr}$, and $\forall \mathcal{S}\subseteq \mathcal{T}\subset \setA_N$ and $j\in\setA_N\backslash \mathcal{T}$, 
\begin{equation*}
\begin{array}{ll}
& \stk( \{ b_i\times ctr_i \}_{i\in \mathcal{S}\cup\{j\}} ) - \stk( \{ b_i\times ctr_i \}_{i\in \mathcal{S}} )  \\
\geq & \stk( \{ b_i\times ctr_i \}_{i\in \mathcal{T}\cup\{j\}} ) - \stk( \{ b_i\times ctr_i \}_{i\in \mathcal{T}} ).
\end{array}
\end{equation*}
Then, we take expectation over the distribution of $\bs{ctr}$ on both sides of the above inequality and finish the proof,
\begin{equation*}
\begin{array}{ll}
& \E_{\bs{ctr}} [ \stk( \{ b_i\times ctr_i \}_{i\in \mathcal{S}\cup\{j\}} ) ] \\ & - \E_{\bs{ctr}} [ \stk( \{ b_i\times ctr_i \}_{i\in \mathcal{S}} ) ] \\
\geq & \E_{\bs{ctr}} [ \stk( \{ b_i\times ctr_i \}_{i\in \mathcal{T}\cup\{j\}} ) ] \\ & - \E_{\bs{ctr}} [ \stk( \{ b_i\times ctr_i \}_{i\in \mathcal{T}} ) ].
\end{array}
\end{equation*}
\end{proof}


\begin{proof}[Proof for Lemma \ref{lem:relation-y-topk}]
For any pair of permutations $(\pi, \pi^\prime)$ that satisfy 
(1) $\pi(i) = \pi^\prime(j) \leq K$; 
(2) $\pi(j) = \pi^\prime(i) > K$; and 
(3) $\pi(k)=\pi^\prime(k)$ for all $k\neq i \wedge k\neq j$,
we can easily verify that $\Pr[\pi|\bs{y}] > \Pr[\pi^\prime|\bs{y}]$. 
Then, summing up on all these pairs of $(\pi, \pi^\prime)$, we have 
$$\Pr[i \in \setA_N^K \wedge j \notin \setA_N^K] 
> \Pr[j \in \setA_N^K \wedge i \notin \setA_N^K],$$
and then 
\begin{equation*}
\begin{aligned}
& \Pr[i \in \setA_N^K] = 
\Pr[i,j \in \setA_N^K] + \Pr[i \in \setA_N^K \wedge j \notin \setA_N^K] 
\\
\geq & \Pr[i,j \in \setA_N^K] + \Pr[j \in \setA_N^K \wedge i \notin \setA_N^K] = \Pr[j \in \setA_N^K].
\end{aligned}
\end{equation*}
\end{proof}

%% file: 21fupa.bbl

\begin{thebibliography}{39}


\ifx \showCODEN    \undefined \def \showCODEN     #1{\unskip}     \fi
\ifx \showDOI      \undefined \def \showDOI       #1{#1}\fi
\ifx \showISBNx    \undefined \def \showISBNx     #1{\unskip}     \fi
\ifx \showISBNxiii \undefined \def \showISBNxiii  #1{\unskip}     \fi
\ifx \showISSN     \undefined \def \showISSN      #1{\unskip}     \fi
\ifx \showLCCN     \undefined \def \showLCCN      #1{\unskip}     \fi
\ifx \shownote     \undefined \def \shownote      #1{#1}          \fi
\ifx \showarticletitle \undefined \def \showarticletitle #1{#1}   \fi
\ifx \showURL      \undefined \def \showURL       {\relax}        \fi
\providecommand\bibfield[2]{#2}
\providecommand\bibinfo[2]{#2}
\providecommand\natexlab[1]{#1}
\providecommand\showeprint[2][]{arXiv:#2}

\bibitem[\protect\citeauthoryear{Balseiro, Deng, Mao, Mirrokni, and
  Zuo}{Balseiro et~al\mbox{.}}{2021}]%
        {balseiro2021landscape}
\bibfield{author}{\bibinfo{person}{Santiago Balseiro}, \bibinfo{person}{Yuan
  Deng}, \bibinfo{person}{Jieming Mao}, \bibinfo{person}{Vahab Mirrokni}, {and}
  \bibinfo{person}{Song Zuo}.} \bibinfo{year}{2021}\natexlab{}.
\newblock \showarticletitle{The Landscape of Auto-Bidding Auctions: Value
  Versus Utility Maximization}.
\newblock \bibinfo{journal}{\emph{Available at SSRN 3785579}}
  (\bibinfo{year}{2021}).
\newblock


\bibitem[\protect\citeauthoryear{Buchbinder, Feldman, Naor, and
  Schwartz}{Buchbinder et~al\mbox{.}}{2014}]%
        {buchbinder2014submodular}
\bibfield{author}{\bibinfo{person}{Niv Buchbinder}, \bibinfo{person}{Moran
  Feldman}, \bibinfo{person}{Joseph Naor}, {and} \bibinfo{person}{Roy
  Schwartz}.} \bibinfo{year}{2014}\natexlab{}.
\newblock \showarticletitle{Submodular maximization with cardinality
  constraints}. In \bibinfo{booktitle}{\emph{Proceedings of the 25th Annual
  ACM-SIAM Symposium on Discrete Algorithms}}. \bibinfo{pages}{1433--1452}.
\newblock


\bibitem[\protect\citeauthoryear{Cao, Qin, Liu, Tsai, and Li}{Cao
  et~al\mbox{.}}{2007}]%
        {cao2007learning}
\bibfield{author}{\bibinfo{person}{Zhe Cao}, \bibinfo{person}{Tao Qin},
  \bibinfo{person}{Tie-Yan Liu}, \bibinfo{person}{Ming-Feng Tsai}, {and}
  \bibinfo{person}{Hang Li}.} \bibinfo{year}{2007}\natexlab{}.
\newblock \showarticletitle{Learning to rank: from pairwise approach to
  listwise approach}. In \bibinfo{booktitle}{\emph{Proceedings of the 24th
  International Conference on Machine Learning}}. \bibinfo{pages}{129--136}.
\newblock


\bibitem[\protect\citeauthoryear{Chekuri and Quanrud}{Chekuri and
  Quanrud}{2019}]%
        {chekuri2019submodular}
\bibfield{author}{\bibinfo{person}{Chandra Chekuri} {and} \bibinfo{person}{Kent
  Quanrud}.} \bibinfo{year}{2019}\natexlab{}.
\newblock \showarticletitle{Submodular function maximization in parallel via
  the multilinear relaxation}. In \bibinfo{booktitle}{\emph{Proceedings of the
  30th Annual ACM-SIAM Symposium on Discrete Algorithms}}. SIAM,
  \bibinfo{pages}{303--322}.
\newblock


\bibitem[\protect\citeauthoryear{Chen, Beutel, Covington, Jain, Belletti, and
  Chi}{Chen et~al\mbox{.}}{2019}]%
        {chen2019top}
\bibfield{author}{\bibinfo{person}{Minmin Chen}, \bibinfo{person}{Alex Beutel},
  \bibinfo{person}{Paul Covington}, \bibinfo{person}{Sagar Jain},
  \bibinfo{person}{Francois Belletti}, {and} \bibinfo{person}{Ed~H Chi}.}
  \bibinfo{year}{2019}\natexlab{}.
\newblock \showarticletitle{Top-k off-policy correction for a REINFORCE
  recommender system}. In \bibinfo{booktitle}{\emph{Proceedings of the 12th ACM
  International Conference on Web Search and Data Mining}}.
  \bibinfo{pages}{456--464}.
\newblock


\bibitem[\protect\citeauthoryear{Cheng, Koc, Harmsen, Shaked, Chandra, Aradhye,
  Anderson, Corrado, Chai, Ispir, et~al\mbox{.}}{Cheng et~al\mbox{.}}{2016}]%
        {cheng2016wide}
\bibfield{author}{\bibinfo{person}{Heng-Tze Cheng}, \bibinfo{person}{Levent
  Koc}, \bibinfo{person}{Jeremiah Harmsen}, \bibinfo{person}{Tal Shaked},
  \bibinfo{person}{Tushar Chandra}, \bibinfo{person}{Hrishi Aradhye},
  \bibinfo{person}{Glen Anderson}, \bibinfo{person}{Greg Corrado},
  \bibinfo{person}{Wei Chai}, \bibinfo{person}{Mustafa Ispir}, {et~al\mbox{.}}}
  \bibinfo{year}{2016}\natexlab{}.
\newblock \showarticletitle{Wide \& deep learning for recommender systems}. In
  \bibinfo{booktitle}{\emph{Proceedings of the 1st Workshop on Deep Learning
  for Recommender Systems}}. \bibinfo{pages}{7--10}.
\newblock


\bibitem[\protect\citeauthoryear{Covington, Adams, and Sargin}{Covington
  et~al\mbox{.}}{2016}]%
        {covington2016deep}
\bibfield{author}{\bibinfo{person}{Paul Covington}, \bibinfo{person}{Jay
  Adams}, {and} \bibinfo{person}{Emre Sargin}.}
  \bibinfo{year}{2016}\natexlab{}.
\newblock \showarticletitle{Deep neural networks for youtube recommendations}.
  In \bibinfo{booktitle}{\emph{Proceedings of the 10th ACM Conference on
  Recommender Systems}}. \bibinfo{pages}{191--198}.
\newblock


\bibitem[\protect\citeauthoryear{Deng, Wang, Tan, Xu, and Gai}{Deng
  et~al\mbox{.}}{2020b}]%
        {deng2020calibrating}
\bibfield{author}{\bibinfo{person}{Chao Deng}, \bibinfo{person}{Hao Wang},
  \bibinfo{person}{Qing Tan}, \bibinfo{person}{Jian Xu}, {and}
  \bibinfo{person}{Kun Gai}.} \bibinfo{year}{2020}\natexlab{b}.
\newblock \showarticletitle{Calibrating user response predictions in online
  advertising}. In \bibinfo{booktitle}{\emph{Joint European Conference on
  Machine Learning and Knowledge Discovery in Databases}}.
  \bibinfo{pages}{208--223}.
\newblock


\bibitem[\protect\citeauthoryear{Deng and Lahaie}{Deng and Lahaie}{2019}]%
        {deng2019testing}
\bibfield{author}{\bibinfo{person}{Yuan Deng} {and}
  \bibinfo{person}{S{\'e}bastien Lahaie}.} \bibinfo{year}{2019}\natexlab{}.
\newblock \showarticletitle{Testing dynamic incentive compatibility in display
  ad auctions}. In \bibinfo{booktitle}{\emph{Proceedings of the 25th ACM SIGKDD
  Conference on Knowledge Discovery \& Data Mining}}.
  \bibinfo{pages}{1616--1624}.
\newblock


\bibitem[\protect\citeauthoryear{Deng, Lahaie, Mirrokni, and Zuo}{Deng
  et~al\mbox{.}}{2020a}]%
        {deng2020data}
\bibfield{author}{\bibinfo{person}{Yuan Deng}, \bibinfo{person}{S{\'e}bastien
  Lahaie}, \bibinfo{person}{Vahab Mirrokni}, {and} \bibinfo{person}{Song Zuo}.}
  \bibinfo{year}{2020}\natexlab{a}.
\newblock \showarticletitle{A data-driven metric of incentive compatibility}.
  In \bibinfo{booktitle}{\emph{Proceedings of the Web Conference 2020}}.
  \bibinfo{pages}{1796--1806}.
\newblock


\bibitem[\protect\citeauthoryear{D{\"u}tting, Feng, Narasimhan, Parkes, and
  Ravindranath}{D{\"u}tting et~al\mbox{.}}{2019}]%
        {dutting2019optimal}
\bibfield{author}{\bibinfo{person}{Paul D{\"u}tting}, \bibinfo{person}{Zhe
  Feng}, \bibinfo{person}{Harikrishna Narasimhan}, \bibinfo{person}{David
  Parkes}, {and} \bibinfo{person}{Sai~Srivatsa Ravindranath}.}
  \bibinfo{year}{2019}\natexlab{}.
\newblock \showarticletitle{Optimal Auctions through Deep Learning}. In
  \bibinfo{booktitle}{\emph{Proceedings of the 36th International Conference on
  Machine Learning}}. \bibinfo{pages}{1706--1715}.
\newblock


\bibitem[\protect\citeauthoryear{Edelman, Ostrovsky, and Schwarz}{Edelman
  et~al\mbox{.}}{2007}]%
        {edelman2007internet}
\bibfield{author}{\bibinfo{person}{Benjamin Edelman}, \bibinfo{person}{Michael
  Ostrovsky}, {and} \bibinfo{person}{Michael Schwarz}.}
  \bibinfo{year}{2007}\natexlab{}.
\newblock \showarticletitle{Internet advertising and the generalized
  second-price auction: Selling billions of dollars worth of keywords}.
\newblock \bibinfo{journal}{\emph{American Economic Review}}
  \bibinfo{volume}{97}, \bibinfo{number}{1} (\bibinfo{year}{2007}),
  \bibinfo{pages}{242--259}.
\newblock


\bibitem[\protect\citeauthoryear{Eksombatchai, Jindal, Liu, Liu, Sharma,
  Sugnet, Ulrich, and Leskovec}{Eksombatchai et~al\mbox{.}}{2018}]%
        {eksombatchai2018pixie}
\bibfield{author}{\bibinfo{person}{Chantat Eksombatchai},
  \bibinfo{person}{Pranav Jindal}, \bibinfo{person}{Jerry~Zitao Liu},
  \bibinfo{person}{Yuchen Liu}, \bibinfo{person}{Rahul Sharma},
  \bibinfo{person}{Charles Sugnet}, \bibinfo{person}{Mark Ulrich}, {and}
  \bibinfo{person}{Jure Leskovec}.} \bibinfo{year}{2018}\natexlab{}.
\newblock \showarticletitle{Pixie: A system for recommending 3+ billion items
  to 200+ million users in real-time}. In \bibinfo{booktitle}{\emph{Proceedings
  of the Web Conference 2018}}. \bibinfo{pages}{1775--1784}.
\newblock


\bibitem[\protect\citeauthoryear{Golrezaei, Lin, Mirrokni, and
  Nazerzadeh}{Golrezaei et~al\mbox{.}}{2021}]%
        {golrezaei2017boosted}
\bibfield{author}{\bibinfo{person}{Negin Golrezaei}, \bibinfo{person}{Max Lin},
  \bibinfo{person}{Vahab Mirrokni}, {and} \bibinfo{person}{Hamid Nazerzadeh}.}
  \bibinfo{year}{2021}\natexlab{}.
\newblock \showarticletitle{Boosted Second Price Auctions: Revenue Optimization
  for Heterogeneous Bidders}. In \bibinfo{booktitle}{\emph{Proceedings of the
  27th ACM SIGKDD Conference on Knowledge Discovery \& Data Mining}}.
  \bibinfo{pages}{447--457}.
\newblock


\bibitem[\protect\citeauthoryear{Guiver and Snelson}{Guiver and
  Snelson}{2009}]%
        {guiver2009bayesian}
\bibfield{author}{\bibinfo{person}{John Guiver} {and} \bibinfo{person}{Edward
  Snelson}.} \bibinfo{year}{2009}\natexlab{}.
\newblock \showarticletitle{Bayesian inference for Plackett-Luce ranking
  models}. In \bibinfo{booktitle}{\emph{Proceedings of the 26th International
  Conference on Machine Learning}}. \bibinfo{pages}{377--384}.
\newblock


\bibitem[\protect\citeauthoryear{He and McAuley}{He and McAuley}{2016}]%
        {he2016ups}
\bibfield{author}{\bibinfo{person}{Ruining He} {and} \bibinfo{person}{Julian
  McAuley}.} \bibinfo{year}{2016}\natexlab{}.
\newblock \showarticletitle{Ups and downs: Modeling the visual evolution of
  fashion trends with one-class collaborative filtering}. In
  \bibinfo{booktitle}{\emph{Proceedings of the Web Conference 2016}}.
  \bibinfo{pages}{507--517}.
\newblock


\bibitem[\protect\citeauthoryear{He, Pan, Jin, Xu, Liu, Xu, Shi, Atallah,
  Herbrich, Bowers, et~al\mbox{.}}{He et~al\mbox{.}}{2014}]%
        {he2014practical}
\bibfield{author}{\bibinfo{person}{Xinran He}, \bibinfo{person}{Junfeng Pan},
  \bibinfo{person}{Ou Jin}, \bibinfo{person}{Tianbing Xu}, \bibinfo{person}{Bo
  Liu}, \bibinfo{person}{Tao Xu}, \bibinfo{person}{Yanxin Shi},
  \bibinfo{person}{Antoine Atallah}, \bibinfo{person}{Ralf Herbrich},
  \bibinfo{person}{Stuart Bowers}, {et~al\mbox{.}}}
  \bibinfo{year}{2014}\natexlab{}.
\newblock \showarticletitle{Practical lessons from predicting clicks on ads at
  facebook}. In \bibinfo{booktitle}{\emph{Proceedings of the 8th International
  Workshop on Data Mining for Online Advertising}}. \bibinfo{pages}{1--9}.
\newblock


\bibitem[\protect\citeauthoryear{Hron, Krauth, Jordan, and Kilbertus}{Hron
  et~al\mbox{.}}{2021}]%
        {hron2021component}
\bibfield{author}{\bibinfo{person}{Jiri Hron}, \bibinfo{person}{Karl Krauth},
  \bibinfo{person}{Michael~I Jordan}, {and} \bibinfo{person}{Niki Kilbertus}.}
  \bibinfo{year}{2021}\natexlab{}.
\newblock \showarticletitle{On component interactions in two-stage recommender
  systems}.
\newblock \bibinfo{journal}{\emph{arXiv preprint arXiv:2106.14979}}
  (\bibinfo{year}{2021}).
\newblock


\bibitem[\protect\citeauthoryear{Kang and McAuley}{Kang and McAuley}{2019}]%
        {kang2019candidate}
\bibfield{author}{\bibinfo{person}{Wang-Cheng Kang} {and}
  \bibinfo{person}{Julian McAuley}.} \bibinfo{year}{2019}\natexlab{}.
\newblock \showarticletitle{Candidate generation with binary codes for
  large-scale top-n recommendation}. In \bibinfo{booktitle}{\emph{Proceedings
  of the 28th ACM international conference on information and knowledge
  management}}. \bibinfo{pages}{1523--1532}.
\newblock


\bibitem[\protect\citeauthoryear{Kleinberg and Raghu}{Kleinberg and
  Raghu}{2018}]%
        {kleinberg2018team}
\bibfield{author}{\bibinfo{person}{Jon Kleinberg} {and}
  \bibinfo{person}{Maithra Raghu}.} \bibinfo{year}{2018}\natexlab{}.
\newblock \showarticletitle{Team performance with test scores}.
\newblock \bibinfo{journal}{\emph{ACM Transactions on Economics and Computation
  (TEAC)}} \bibinfo{volume}{6}, \bibinfo{number}{3-4} (\bibinfo{year}{2018}),
  \bibinfo{pages}{1--26}.
\newblock


\bibitem[\protect\citeauthoryear{Lahaie and Pennock}{Lahaie and
  Pennock}{2007}]%
        {lahaie2007revenue}
\bibfield{author}{\bibinfo{person}{S{\'e}bastien Lahaie} {and}
  \bibinfo{person}{David~M Pennock}.} \bibinfo{year}{2007}\natexlab{}.
\newblock \showarticletitle{Revenue analysis of a family of ranking rules for
  keyword auctions}. In \bibinfo{booktitle}{\emph{Proceedings of the 8th ACM
  Conference on Electronic Commerce}}. \bibinfo{pages}{50--56}.
\newblock


\bibitem[\protect\citeauthoryear{Lee, Orten, Dasdan, and Li}{Lee
  et~al\mbox{.}}{2012}]%
        {lee2012estimating}
\bibfield{author}{\bibinfo{person}{Kuang-chih Lee}, \bibinfo{person}{Burkay
  Orten}, \bibinfo{person}{Ali Dasdan}, {and} \bibinfo{person}{Wentong Li}.}
  \bibinfo{year}{2012}\natexlab{}.
\newblock \showarticletitle{Estimating conversion rate in display advertising
  from past erformance data}. In \bibinfo{booktitle}{\emph{Proceedings of the
  18th ACM SIGKDD International Conference on Knowledge Discovery and Data
  Mining}}. \bibinfo{pages}{768--776}.
\newblock


\bibitem[\protect\citeauthoryear{Liu, Yu, Zhang, Zheng, Rong, Lv, Huo, Wang,
  Chen, Xu, Wu, Chen, and Zhu}{Liu et~al\mbox{.}}{2021}]%
        {liu2021neural}
\bibfield{author}{\bibinfo{person}{Xiangyu Liu}, \bibinfo{person}{Chuan Yu},
  \bibinfo{person}{Zhilin Zhang}, \bibinfo{person}{Zhenzhe Zheng},
  \bibinfo{person}{Yu Rong}, \bibinfo{person}{Hongtao Lv}, \bibinfo{person}{Da
  Huo}, \bibinfo{person}{Yiqing Wang}, \bibinfo{person}{Dagui Chen},
  \bibinfo{person}{Jian Xu}, \bibinfo{person}{Fan Wu}, \bibinfo{person}{Guihai
  Chen}, {and} \bibinfo{person}{Xiaoqiang Zhu}.}
  \bibinfo{year}{2021}\natexlab{}.
\newblock \showarticletitle{Neural auction: End-to-end learning of auction
  mechanisms for e-commerce advertising}. In
  \bibinfo{booktitle}{\emph{Proceedings of the 27th ACM SIGKDD Conference on
  Knowledge Discovery \&; Data Mining}}. \bibinfo{pages}{3354–3364}.
\newblock


\bibitem[\protect\citeauthoryear{Ma, Zhao, Yi, Yang, Chen, Tang, Hong, and
  Chi}{Ma et~al\mbox{.}}{2020}]%
        {ma2020off}
\bibfield{author}{\bibinfo{person}{Jiaqi Ma}, \bibinfo{person}{Zhe Zhao},
  \bibinfo{person}{Xinyang Yi}, \bibinfo{person}{Ji Yang},
  \bibinfo{person}{Minmin Chen}, \bibinfo{person}{Jiaxi Tang},
  \bibinfo{person}{Lichan Hong}, {and} \bibinfo{person}{Ed~H Chi}.}
  \bibinfo{year}{2020}\natexlab{}.
\newblock \showarticletitle{Off-policy learning in two-stage recommender
  systems}. In \bibinfo{booktitle}{\emph{Proceedings of the Web Conference
  2020}}. \bibinfo{pages}{463--473}.
\newblock


\bibitem[\protect\citeauthoryear{McAuley, Targett, Shi, and Van
  Den~Hengel}{McAuley et~al\mbox{.}}{2015}]%
        {mcauley2015image}
\bibfield{author}{\bibinfo{person}{Julian McAuley},
  \bibinfo{person}{Christopher Targett}, \bibinfo{person}{Qinfeng Shi}, {and}
  \bibinfo{person}{Anton Van Den~Hengel}.} \bibinfo{year}{2015}\natexlab{}.
\newblock \showarticletitle{Image-based recommendations on styles and
  substitutes}. In \bibinfo{booktitle}{\emph{Proceedings of the 38th
  international ACM SIGIR Conference on Research and Development in Information
  Retrieval}}. \bibinfo{pages}{43--52}.
\newblock


\bibitem[\protect\citeauthoryear{Mehta, Nadav, Psomas, and Rubinstein}{Mehta
  et~al\mbox{.}}{2020}]%
        {mehta2020hitting}
\bibfield{author}{\bibinfo{person}{Aranyak Mehta}, \bibinfo{person}{Uri Nadav},
  \bibinfo{person}{Alexandros Psomas}, {and} \bibinfo{person}{Aviad
  Rubinstein}.} \bibinfo{year}{2020}\natexlab{}.
\newblock \showarticletitle{Hitting the high notes: Subset selection for
  maximizing expected order statistics}.
\newblock \bibinfo{journal}{\emph{Advances in Neural Information Processing
  Systems}}  \bibinfo{volume}{33} (\bibinfo{year}{2020}).
\newblock


\bibitem[\protect\citeauthoryear{Myerson}{Myerson}{1981}]%
        {myerson1981optimal}
\bibfield{author}{\bibinfo{person}{Roger~B Myerson}.}
  \bibinfo{year}{1981}\natexlab{}.
\newblock \showarticletitle{Optimal auction design}.
\newblock \bibinfo{journal}{\emph{Mathematics of Operations Research}}
  \bibinfo{volume}{6}, \bibinfo{number}{1} (\bibinfo{year}{1981}),
  \bibinfo{pages}{58--73}.
\newblock


\bibitem[\protect\citeauthoryear{Nemhauser, Wolsey, and Fisher}{Nemhauser
  et~al\mbox{.}}{1978}]%
        {nemhauser1978analysis}
\bibfield{author}{\bibinfo{person}{George~L Nemhauser},
  \bibinfo{person}{Laurence~A Wolsey}, {and} \bibinfo{person}{Marshall~L
  Fisher}.} \bibinfo{year}{1978}\natexlab{}.
\newblock \showarticletitle{An analysis of approximations for maximizing
  submodular set functions—I}.
\newblock \bibinfo{journal}{\emph{Mathematical Programming}}
  \bibinfo{volume}{14}, \bibinfo{number}{1} (\bibinfo{year}{1978}),
  \bibinfo{pages}{265--294}.
\newblock


\bibitem[\protect\citeauthoryear{Shen, Peng, Liu, Zhang, Qian, Hong, Guo, Ding,
  Lu, and Tang}{Shen et~al\mbox{.}}{2020}]%
        {shen2020reinforcement}
\bibfield{author}{\bibinfo{person}{Weiran Shen}, \bibinfo{person}{Binghui
  Peng}, \bibinfo{person}{Hanpeng Liu}, \bibinfo{person}{Michael Zhang},
  \bibinfo{person}{Ruohan Qian}, \bibinfo{person}{Yan Hong},
  \bibinfo{person}{Zhi Guo}, \bibinfo{person}{Zongyao Ding},
  \bibinfo{person}{Pengjun Lu}, {and} \bibinfo{person}{Pingzhong Tang}.}
  \bibinfo{year}{2020}\natexlab{}.
\newblock \showarticletitle{Reinforcement mechanism design: With applications
  to dynamic pricing in sponsored search auctions}. In
  \bibinfo{booktitle}{\emph{Proceedings of the AAAI Conference on Artificial
  Intelligence}}, Vol.~\bibinfo{volume}{34}. \bibinfo{pages}{2236--2243}.
\newblock


\bibitem[\protect\citeauthoryear{Varian}{Varian}{2007}]%
        {varian2007position}
\bibfield{author}{\bibinfo{person}{Hal~R Varian}.}
  \bibinfo{year}{2007}\natexlab{}.
\newblock \showarticletitle{Position auctions}.
\newblock \bibinfo{journal}{\emph{international Journal of industrial
  Organization}} \bibinfo{volume}{25}, \bibinfo{number}{6}
  (\bibinfo{year}{2007}), \bibinfo{pages}{1163--1178}.
\newblock


\bibitem[\protect\citeauthoryear{Varian and Harris}{Varian and Harris}{2014}]%
        {varian2014vcg}
\bibfield{author}{\bibinfo{person}{Hal~R Varian} {and}
  \bibinfo{person}{Christopher Harris}.} \bibinfo{year}{2014}\natexlab{}.
\newblock \showarticletitle{The VCG auction in theory and practice}.
\newblock \bibinfo{journal}{\emph{American Economic Review}}
  \bibinfo{volume}{104}, \bibinfo{number}{5} (\bibinfo{year}{2014}),
  \bibinfo{pages}{442--45}.
\newblock


\bibitem[\protect\citeauthoryear{Wang, Zhao, Jiang, Zhou, Zhu, and Gai}{Wang
  et~al\mbox{.}}{2020}]%
        {wang2020cold}
\bibfield{author}{\bibinfo{person}{Zhe Wang}, \bibinfo{person}{Liqin Zhao},
  \bibinfo{person}{Biye Jiang}, \bibinfo{person}{Guorui Zhou},
  \bibinfo{person}{Xiaoqiang Zhu}, {and} \bibinfo{person}{Kun Gai}.}
  \bibinfo{year}{2020}\natexlab{}.
\newblock \showarticletitle{COLD: Towards the Next Generation of Pre-Ranking
  System}.
\newblock \bibinfo{journal}{\emph{arXiv preprint arXiv:2007.16122}}
  (\bibinfo{year}{2020}).
\newblock


\bibitem[\protect\citeauthoryear{Wilkens, Cavallo, and Niazadeh}{Wilkens
  et~al\mbox{.}}{2017}]%
        {wilkens2017gsp}
\bibfield{author}{\bibinfo{person}{Christopher~A Wilkens},
  \bibinfo{person}{Ruggiero Cavallo}, {and} \bibinfo{person}{Rad Niazadeh}.}
  \bibinfo{year}{2017}\natexlab{}.
\newblock \showarticletitle{GSP: the cinderella of mechanism design}. In
  \bibinfo{booktitle}{\emph{Proceedings of the Web Conference 2017}}.
  \bibinfo{pages}{25--32}.
\newblock


\bibitem[\protect\citeauthoryear{Wu, Chen, Yang, Wang, Tan, Zhang, Xu, and
  Gai}{Wu et~al\mbox{.}}{2018}]%
        {wu2018budget}
\bibfield{author}{\bibinfo{person}{Di Wu}, \bibinfo{person}{Xiujun Chen},
  \bibinfo{person}{Xun Yang}, \bibinfo{person}{Hao Wang}, \bibinfo{person}{Qing
  Tan}, \bibinfo{person}{Xiaoxun Zhang}, \bibinfo{person}{Jian Xu}, {and}
  \bibinfo{person}{Kun Gai}.} \bibinfo{year}{2018}\natexlab{}.
\newblock \showarticletitle{Budget constrained bidding by model-free
  reinforcement learning in display advertising}. In
  \bibinfo{booktitle}{\emph{Proceedings of the 27th ACM International
  Conference on Information and Knowledge Management}}.
  \bibinfo{pages}{1443--1451}.
\newblock


\bibitem[\protect\citeauthoryear{Yang, Li, Wang, Wu, Tan, Xu, and Gai}{Yang
  et~al\mbox{.}}{2019}]%
        {yang2019bid}
\bibfield{author}{\bibinfo{person}{Xun Yang}, \bibinfo{person}{Yasong Li},
  \bibinfo{person}{Hao Wang}, \bibinfo{person}{Di Wu}, \bibinfo{person}{Qing
  Tan}, \bibinfo{person}{Jian Xu}, {and} \bibinfo{person}{Kun Gai}.}
  \bibinfo{year}{2019}\natexlab{}.
\newblock \showarticletitle{Bid optimization by multivariable control in
  display advertising}. In \bibinfo{booktitle}{\emph{Proceedings of the 25th
  ACM SIGKDD Conference on Knowledge Discovery \& Data Mining}}.
  \bibinfo{pages}{1966--1974}.
\newblock


\bibitem[\protect\citeauthoryear{Yi, Yang, Hong, Cheng, Heldt, Kumthekar, Zhao,
  Wei, and Chi}{Yi et~al\mbox{.}}{2019}]%
        {yi2019sampling}
\bibfield{author}{\bibinfo{person}{Xinyang Yi}, \bibinfo{person}{Ji Yang},
  \bibinfo{person}{Lichan Hong}, \bibinfo{person}{Derek~Zhiyuan Cheng},
  \bibinfo{person}{Lukasz Heldt}, \bibinfo{person}{Aditee Kumthekar},
  \bibinfo{person}{Zhe Zhao}, \bibinfo{person}{Li Wei}, {and}
  \bibinfo{person}{Ed Chi}.} \bibinfo{year}{2019}\natexlab{}.
\newblock \showarticletitle{Sampling-bias-corrected neural modeling for large
  corpus item recommendations}. In \bibinfo{booktitle}{\emph{Proceedings of the
  13th ACM Conference on Recommender Systems}}. \bibinfo{pages}{269--277}.
\newblock


\bibitem[\protect\citeauthoryear{Zhou, Mou, Fan, Pi, Bian, Zhou, Zhu, and
  Gai}{Zhou et~al\mbox{.}}{2019}]%
        {zhou2019deep}
\bibfield{author}{\bibinfo{person}{Guorui Zhou}, \bibinfo{person}{Na Mou},
  \bibinfo{person}{Ying Fan}, \bibinfo{person}{Qi Pi}, \bibinfo{person}{Weijie
  Bian}, \bibinfo{person}{Chang Zhou}, \bibinfo{person}{Xiaoqiang Zhu}, {and}
  \bibinfo{person}{Kun Gai}.} \bibinfo{year}{2019}\natexlab{}.
\newblock \showarticletitle{Deep interest evolution network for click-through
  rate prediction}. In \bibinfo{booktitle}{\emph{Proceedings of the AAAI
  Conference on Artificial Intelligence}}, Vol.~\bibinfo{volume}{33}.
  \bibinfo{pages}{5941--5948}.
\newblock


\bibitem[\protect\citeauthoryear{Zhou, Zhu, Song, Fan, Zhu, Ma, Yan, Jin, Li,
  and Gai}{Zhou et~al\mbox{.}}{2018}]%
        {zhou2018deep}
\bibfield{author}{\bibinfo{person}{Guorui Zhou}, \bibinfo{person}{Xiaoqiang
  Zhu}, \bibinfo{person}{Chenru Song}, \bibinfo{person}{Ying Fan},
  \bibinfo{person}{Han Zhu}, \bibinfo{person}{Xiao Ma},
  \bibinfo{person}{Yanghui Yan}, \bibinfo{person}{Junqi Jin},
  \bibinfo{person}{Han Li}, {and} \bibinfo{person}{Kun Gai}.}
  \bibinfo{year}{2018}\natexlab{}.
\newblock \showarticletitle{Deep interest network for click-through rate
  prediction}. In \bibinfo{booktitle}{\emph{Proceedings of the 24th ACM SIGKDD
  Conference on Knowledge Discovery \& Data Mining}}.
  \bibinfo{pages}{1059--1068}.
\newblock


\bibitem[\protect\citeauthoryear{Zhu, Jin, Tan, Pan, Zeng, Li, and Gai}{Zhu
  et~al\mbox{.}}{2017}]%
        {zhu2017optimized}
\bibfield{author}{\bibinfo{person}{Han Zhu}, \bibinfo{person}{Junqi Jin},
  \bibinfo{person}{Chang Tan}, \bibinfo{person}{Fei Pan},
  \bibinfo{person}{Yifan Zeng}, \bibinfo{person}{Han Li}, {and}
  \bibinfo{person}{Kun Gai}.} \bibinfo{year}{2017}\natexlab{}.
\newblock \showarticletitle{Optimized cost per click in taobao display
  advertising}. In \bibinfo{booktitle}{\emph{Proceedings of the 23rd ACM SIGKDD
  Conference on Knowledge Discovery \& Data Mining}}.
  \bibinfo{pages}{2191--2200}.
\newblock


\end{thebibliography}
